\documentclass[11pt]{article}

\usepackage[margin=1in]{geometry}  
\usepackage{graphicx}              
\usepackage{amsmath}               
\usepackage{amsfonts}              
\usepackage{amsthm}                
\usepackage{color}
\usepackage{caption}
\usepackage{subcaption}

\begin{document}

\nocite{*}

\title{Mediation Analysis for Censored Survival Data under an Accelerated Failure Time Model}
\date{}
\author{Isabel R. Fulcher, Eric Tchetgen Tchetgen, Paige L. Williams}  

\date{}
\maketitle
\vspace{-1.4cm}
\begin{center} Department of Biostatistics, Harvard University \end{center}
\vspace{.1cm}
\maketitle

\begin{abstract}
Recent advances in causal mediation analysis have formalized conditions for estimating direct and indirect effects in various contexts. These approaches have been extended to a number of models for survival outcomes including accelerated failure time (AFT) models which are widely used in a broad range of health applications given their intuitive interpretation. In this setting, it has been suggested that under standard assumptions, the ``difference" and ``product" methods produce equivalent estimates of the indirect effect of exposure on the survival outcome. We formally show that these two methods may produce substantially different estimates in the presence of censoring or truncation, due to a form of model misspecification. Specifically, we establish that while the product method remains valid under standard assumptions in the presence of independent censoring, the difference method can be biased in the presence of such censoring whenever the error distribution of the AFT model fails to be collapsible upon marginalizing over the mediator. This will invariably be the case for most choices of mediator and outcome error distributions. A notable exception arises in case of normal mediator-normal outcome where we show consistency of both difference and product estimators in the presence of independent censoring. These results are confirmed in simulation studies and two data applications. 
\end{abstract}

\section*{Background}
Numerous papers have in recent years laid the foundation for causal mediation analysis in the context of linear and nonlinear models for continuous, binary, and survival outcomes, and likewise in situations where an interaction may be present between the exposure and the mediator.$^{1-7}$ These advances have clarified conditions under which traditional mediation techniques for estimating the indirect effect, such as the ``product" and ``difference" methods, are equivalent.$^{3,5}$ In a recent commentary, VanderWeele$^{8}$ established that the well-known equivalence of the difference and product methods in linear models with no exposure-mediator interaction holds exactly for the mean of the log of a survival outcome under a certain accelerated failure time (AFT) model. This result confirmed previous findings by Tein and MacKinnon$^{9}$ who showed in simulation studies that under a Weibull AFT model the product and difference estimators were consistent for the indirect effect; however, both Tein and MacKinnon$^{9}$ and VanderWeele$^{8}$ only considered a setting in which censoring was absent. \\

In practice, outcomes in survival data are typically subject to some form of censoring, primarily due to loss to follow-up and administrative censoring. The AFT model is a prominent approach for handling censored survival data in the health sciences and has become widely used in practice by epidemiologists partly because it is readily available from popular standard commercial software packages such as PROC LIFEREG in SAS,$^{10}$ streg in Stata,$^{11}$ and survreg in R.$^{12}$ Another approach to handling censored survival data is the Cox proportional hazards model, but the regression coefficients from these models can be hard to interpret due to built-in selection bias as they condition on information about the outcome.$^{13,14}$ The central role of survival models in medical research coupled with the increasing popularity of causal mediation analyses has led to method development and software that will estimate direct and indirect effects. Recently, Valeri and Vanderweele$^{15}$ created a SAS macro to estimate direct and indirect effects in censored survival data using AFT and Cox PH models. They cautioned that the estimated direct and indirect effects for a Cox PH model are based on a rare disease assumption, while the AFT model does not require such an assumption.$^{15}$ In a more recent paper, Gelfand et al.$^{16}$ encourage the use of AFT models over Cox PH models in mediation analysis. A commonly noted drawback of the AFT model is that the event times are assumed to follow a specific distribution, but Gelfand et al.$^{16}$ argue that the breadth of distributions available can capture the variability in survival data, and the Weibull distribution can represent distributions commonly found in clinical research.\\ 

For mediation analysis with a censored survival outcome and assuming no exposure-mediator interaction in the outcome model, researchers can easily estimate direct and indirect effects in standard software through a combination of AFT and linear models (see R Code in appendix). In doing so, researchers may estimate the indirect effect either by the product or difference method, as the results by Tein and MacKinnon$^{9}$ and VanderWeele$^{8}$ may inadvertently lead one to incorrectly assume that both estimators are valid. However, Gelfand et al.$^{16}$ demonstrate in simulation studies that in the presence of right censoring the product and difference estimators are not necessarily equivalent under a Weibull AFT model. Interestingly, they note that the product method appeared to remain unaffected by right censoring, while the difference method underestimated the indirect effect in their simulation studies.$^{16}$ The results of our paper will provide the theoretical underpinning for the conclusions drawn from simulation studies by Tein and MacKinnon$^{9}$ and Gelfand et al.$^{16}$. We supply formal justification for their conjecture that the difference method will often fail to provide a consistent estimator of the indirect effect under an AFT model even when the product method does.  \\

Specifically, in this paper, we formally establish that the equivalence between the product and difference method generally fails in the presence of censoring, primarily due to lack of consistency of the difference method arising from a form of model misspecification. We will formally show that this form of model misspecification gives rise to bias in the difference method estimator when censoring is present. However, in the absence of censoring, this form of model misspecification is relatively benign and does not generally induce bias in the estimated indirect effect using the difference method. This misspecification does not arise in the special case of normal mediator-normal outcome model, and, thus, both difference and product estimators are consistent in the presence of independent censoring. In a simulation study, we confirm these results for normal and Weibull distributed time-to-event outcomes, respectively. We also consider the implication of our findings in estimating the indirect effect of HIV status mediated by height for age at sexual maturity in the Pediatric HIV/AIDS Cohort Study (PHACS) and Pediatric AIDS Clinical Trials Group 219C (PACTG) studies and the indirect effect of combination treatment mediated by viral suppression on time to death or opportunistic infection among HIV-infected adults using multiple studies of the AIDS Clinical Trial Group.$^{17-20}$ Although the paper focuses on the implications of censoring, all of the main results  hold for left truncation as we formally show in the appendix. \\

\subsection*{Notation and Assumptions}
Throughout, we focus on a binary exposure $A$, continuous mediator $M$, and time-to-event outcome $T$. To simplify the presentation, we do not explicitly include pre-exposure covariates, and therefore, for all practical purposes, our analysis may be viewed as if we had conditioned on a specific level of such covariates. However, in the appendix formal statements of our main results and corresponding proofs explicitly account for covariates. Let $M(a)$ denote the counterfactual mediator had the exposure taken value $a$ and $T(a) = T(a, M(a))$ denote the counterfactual outcome had exposure taken value $a$. In mediation analysis, we will also consider the counterfactual outcome $T(a, M(a^*))$ had exposure taken value $a=0,1$ and the mediator taken the value it would have under $a^*=0,1$. \\ 

\noindent We consider the following models for the survival outcome $T$ and mediator $M$: 

 		\begin{equation}
                \log T=\beta_{0}+\beta_{a}A+\beta_{m}M+\sigma \varepsilon \label{full model}
                \end{equation}

                \begin{equation}
                M=\alpha_{0}+\alpha_{a}A+\xi  \label{mediator model}
                \end{equation}

\noindent where $\xi$ has mean zero and is independent of $A$, $\varepsilon$ follows a known distribution and is independent of $A$ and $M$, and $\sigma$ is some positive scale parameter. Note that model $\left(  \ref{full model}\right)$ assumes no exposure-mediator interaction, which is necessary for possible equivalence between the product and difference representation of the indirect effect. \\

\noindent As shown in the appendix, under treatment randomization (A1) and cross-world counterfactuals independence (A2) assumptions and following the same reasoning as Pearl,$^2$ the average natural (or pure) direct or indirect effects on the log mean scale is nonparametrically identified. (A1) is a standard no unmeasured confounding assumption of the effects of $A$ on $(M,Y)$, while (A2) is a somewhat stronger no unmeasured confounding assumption of the effects of $M$ on $Y$.$^{21}$ The natural direct ($NDE(a,a^*)$) and indirect ($NIE(a,a^*)$) effects for the log-survival time are defined in terms of these counterfactuals and under models $\left(  \ref{full model}\right)$ and $\left(  \ref{mediator model}\right)$ we have that:

\begin{align*}
NDE\left(  a,a^{\ast}\right)   &  =  E \{  \log  T(a,M(a^{\ast}))  \}
- E \{ \log T(a^{\ast},M(a^{\ast})) \}  =\beta_{a}\left(  a-a^{\ast}\right) \\
NIE\left(  a,a^{\ast}\right)   &  = E \{ \log   T(a,M(a))  \} -
E \{ \log T(a,M(a^{\ast}))  \} =\beta_{m}\alpha_{a}\left(  a-a^{\ast}\right)
\end{align*}

\noindent Letting $a=1$ and $a^*=0$, this leads to a natural direct effect of $\beta_a$ and a natural indirect effect given by the product rule $\beta_m \alpha_a$. The expression for the difference method, given by $\tau_a - \beta_a$, is obtained upon marginalizing over $M$ by positing a second accelerated failure time model for $T$ as a function of $A$ only, which shall be referred to as the reduced form model and is typically specified as followed:

\begin{equation}
                \log T=\beta_{0}^* + \tau_a A+ \widetilde{\sigma} \widetilde{\varepsilon} \label{reduced model}
                \end{equation} 

\noindent where $\tilde{\varepsilon}$ independent of $A$ and typically assumed to follow the same distribution as $\varepsilon$ in $\left(  \ref{full model}\right)$, and $\tilde{\sigma}$ is a positive scale parameter. This specification is also used by both Vanderweele$^8$ and Tein and MacKinnon$^9$. Under this formulation, we see that $\tau_a$ is the total effect of $A$ on the mean of $T$ on the log scale and satisfies: 

		\begin{equation}
                \tau_a = \alpha_a \beta_m + \beta_a  \label{difference rule} 
                \end{equation}

\noindent This equivalence follows from direct substitution of equation  $\left(  \ref{mediator model}\right)$ into $\left(  \ref{full model}\right)$ and evaluation of total effect $\tau_a$. 
 
\subsection*{Equivalence of the Product and Difference Method} 

As we discuss further below, equation $\left(  \ref{difference rule}\right)$ is usually mis-specified because the error distribution specified in models $\left(  \ref{full model}\right)$ and $\left(  \ref{mediator model}\right)$ completely determine the error distribution in model $\left(  \ref{reduced model}\right)$ as a convolution of these two laws (see appendix (A7)). Unless the error distribution of model $\left(  \ref{reduced model}\right)$ is carefully chosen to match this convolution, the model will be mis-specified. This convolution seldom reduces to a standard model typically implemented in off-the-shelf software when $M$ and $T$ follow standard distributions. For instance, suppose that $M$ is assumed normal and $T$ assumed to follow a Weibull distribution, then the reduced form error distribution is a convolution of a normal with a Weibull distribution, which is neither normal nor Weibull and is in fact not of a standard closed form (see appendix (A10)). In this case, the assumption that the error in the reduced model is Weibull is clearly incorrect. A fairly prominent setting in which the reduced form model is correctly specified is the normal mediator-normal outcome model, in which case the error distribution of model $\left(  \ref{reduced model}\right)$ is also normal.  \\

As shown in the appendix, in the absence of censoring, maximum likelihood estimation of the reduced form model will be consistent for the total effect $\tau_a = \alpha_{a}\beta_{m}+ \beta_{a}$ even if the error distribution is mis-specified. This follows from the fact that in the absence of censoring, consistency of the estimated regression parameters in an AFT model depends on correct specification of the regression model, not on the choice of error distribution. In contrast, result (A8) in the appendix formally establishes that in the presence of censoring, maximum likelihood estimation of the reduced form model will fail to be consistent when the error distribution is incorrect because the corresponding score function fails to be unbiased, which is a basic requirement of consistency. Under correct specification of models $\left(  \ref{full model}\right)$ and $\left(  \ref{mediator model}\right)$, the product method corresponds to the maximum likelihood estimator and is guaranteed to be consistent whether or not censoring is present and irrespective of choice of models for residual errors. \\

On these theoretical grounds, we conclude that one should exert caution when using the difference method in the presence of censoring (or as shown in the appendix, in the presence of truncation), as it is prone to model misspecification of model $\left(  \ref{reduced model}\right)$ even when models $\left(  \ref{full model}\right)$ and $\left(  \ref{mediator model}\right)$  are correctly specified. When these two models are correctly specified, the product method gives a valid estimator for the indirect effect. In the next section, we illustrate this phenomenon in extensive simulation studies and two separate applications. \\

\section*{Simulation}

In simulation studies, we considered two scenarios, one where $T$ is normal and the other where $T$ is Weibull distributed. In both settings, $A$ was generated Bernoulli with probability equal to .5. In the first setting, $M$ was generated from a normal model with mean $\alpha_0 + \alpha_aA$ and variance 1, where $(\alpha_0, \alpha_a) = (0,-.5)$. The time-to-event outcome, $T$, was generated from a normal distribution with mean $\beta_0 + \beta_aA + \beta_mM$ and variance 1, where $(\beta_0, \beta_a, \beta_m) = (180,4,-4)$. We investigated the following three censoring scenarios: no censoring, 70\% right-censored and the remaining 30\% with observed event times, and 70\% right-censored and the remaining 30\% interval-censored. Models $\left(  \ref{full model}\right)$, $\left(  \ref{mediator model}\right)$, and $\left(  \ref{reduced model}\right)$ were estimated using survreg in the R survival package$^{12}$, with gaussian time-to-event distribution for models  $\left(  \ref{full model}\right)$ and  $\left(  \ref{reduced model}\right)$ and a linear regression for $\left(  \ref{mediator model}\right)$. For the right-censoring only setting, a censoring distribution was generated from a normal distribution to yield approximately $70\%$ censored. For the right and interval-censoring setting, the same right-censoring distribution was used, and a censoring interval was generated for observed event times. The length of the censoring interval was generated from a multinomial distribution; to choose where on the interval the true time occurred, we generated a proportion from a uniform(0,1) distribution.  \\

\noindent For the second setting with a Weibull distributed time-to-event, the mediator $M$ was generated from a normal model with mean $\alpha_0 + \alpha_aA$ and variance 1, where $(\alpha_0, \alpha_a) = (0,-.3)$. Lastly, the time-to-event outcome was generated as $\beta_0 + \beta_aA + \beta_mM + \sigma \epsilon$, where $(\beta_0, \beta_a, \beta_m) = (4,.5,-.6)$, $\sigma$ is .25, and $\epsilon$ is the extreme value density. For this model, we investigated the following two censoring scenarios: no censoring and 30\% right-censored. Models $\left(  \ref{full model}\right)$, $\left(  \ref{mediator model}\right)$, and $\left(  \ref{reduced model}\right)$ were fit using survreg in the R survival package$^{12}$ with a Weibull distribution for time. For the right censoring scenario, a censoring distribution was generated for a Weibull distribution to yield $\sim 30\%$ censored. We performed 10,000 simulations for each scenario, with sample size ranging from 800 to 4,000. \\

\noindent  We evaluated the following characteristics for each distribution and censoring type: absolute proportion difference between the estimators ($| \widehat{IE}_{p} - \widehat{IE}_{d} |/ \widehat{IE}_{p} $) and proportion bias of each estimator ($| \widehat{IE}_{p} - IE | / IE$ and $|\widehat{IE}_{d} - IE | / \underline{}IE$), where $IE$ is the true indirect effect,  $\widehat{IE}_{p}$ is the Monte Carlo mean of the product estimator, and $\widehat{IE}_{d}$ is the Monte Carlo mean of the difference estimator. Simulation results for each scenario are summarized in Figures 1 and 2. \\

\begin{figure}
\centering
\caption{Normal AFT Simulation, Product vs. Difference Method for Indirect Effect}
\begin{subfigure}{.57\textwidth}
    \includegraphics[scale=.45]{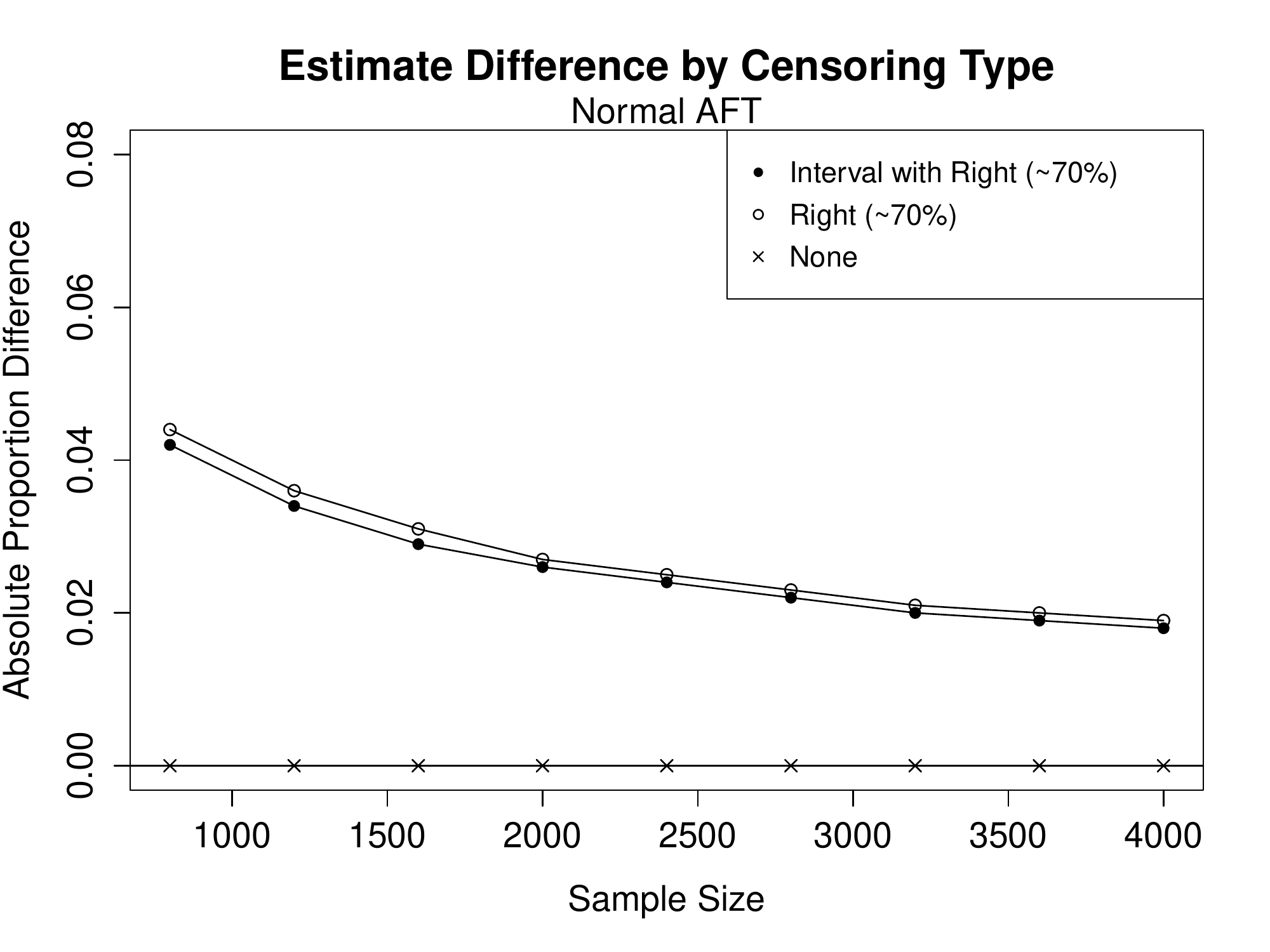}
  \caption{}
\end{subfigure}\\
\begin{subfigure}{.57\textwidth}
  \includegraphics[scale=.45]{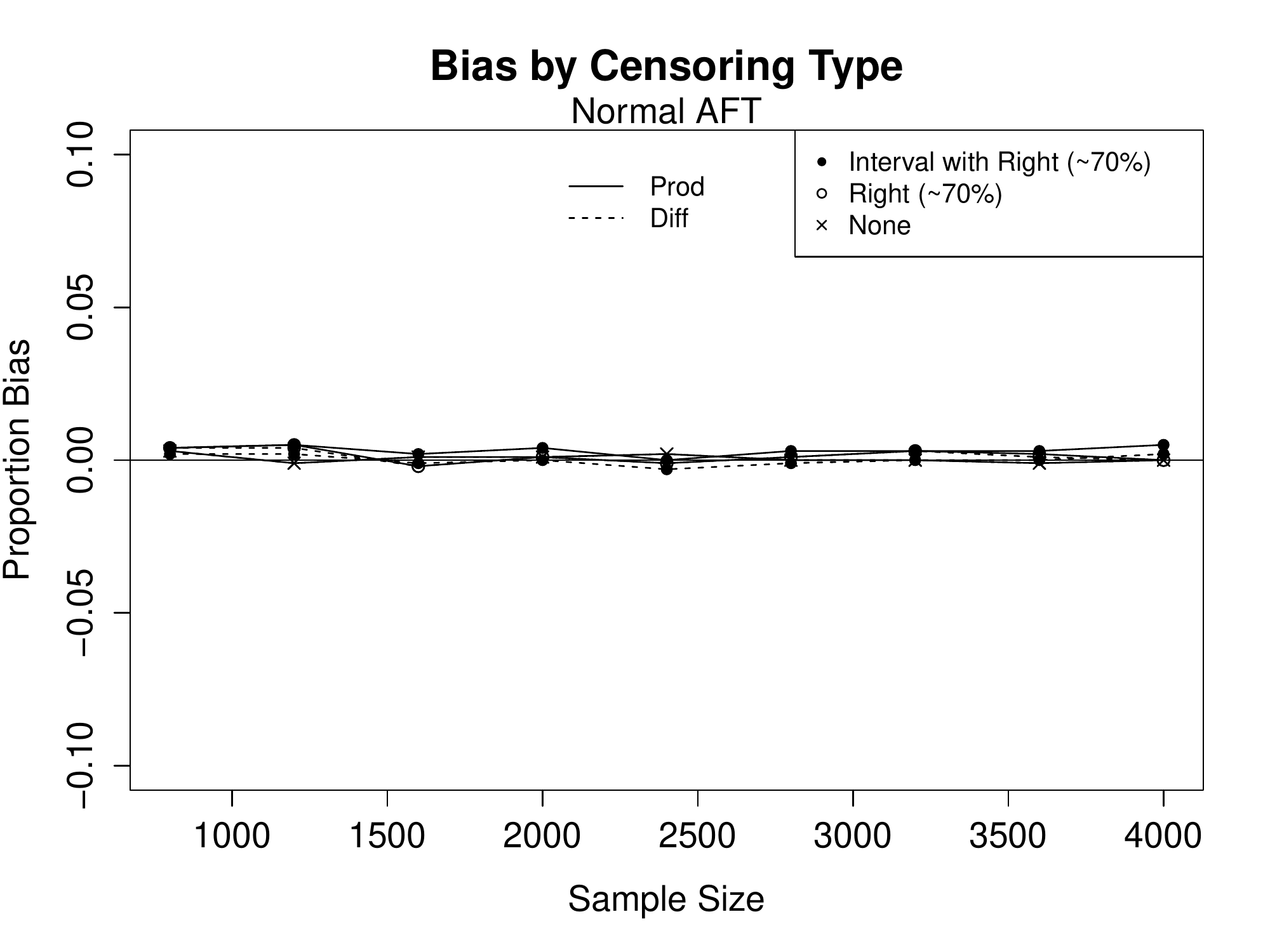}
  \caption{}
\end{subfigure}
\end{figure}

\begin{figure}
\centering
\caption{Weibull AFT Simulation, Product vs. Difference Method for Indirect Effect}
\begin{subfigure}{.57\textwidth}
    \includegraphics[scale=.45]{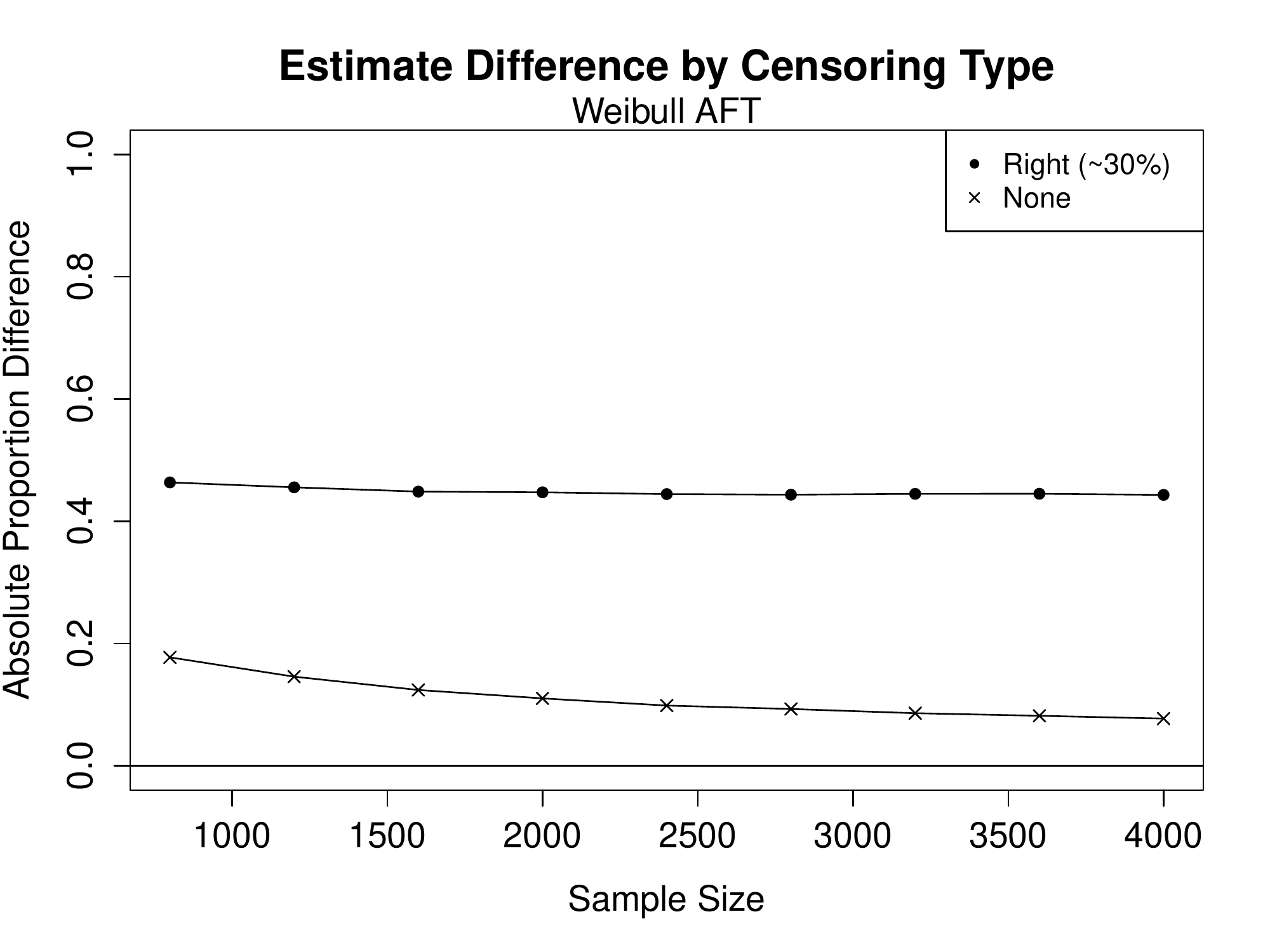}
  \caption{}
\end{subfigure}\\
\begin{subfigure}{.57\textwidth}
  \includegraphics[scale=.45]{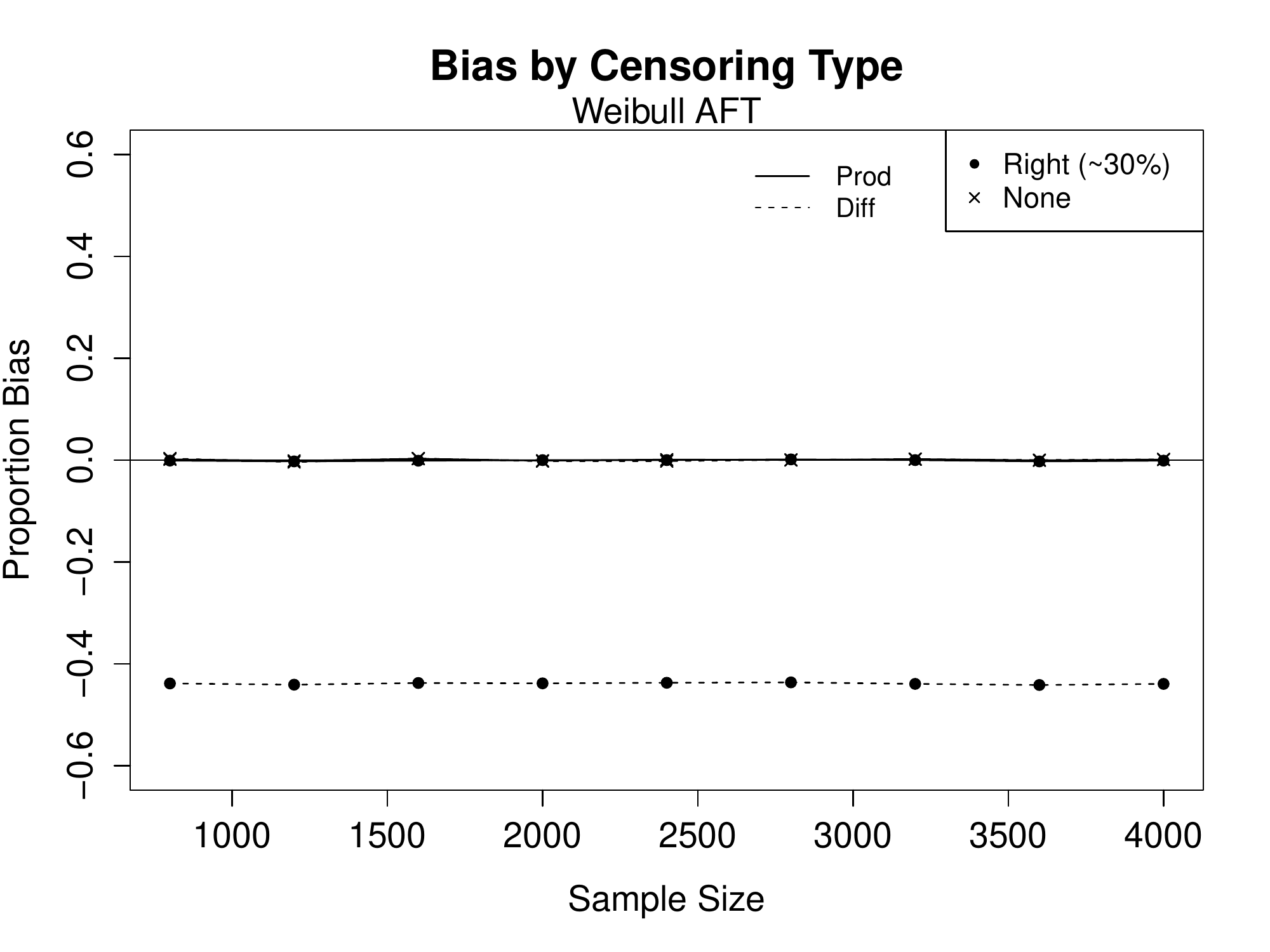}
  \caption{}
\end{subfigure}
\end{figure}
        
\noindent Figure 1a shows the absolute proportion difference between the product and difference method under the normal model. In the absence of censoring, the proportion difference was identically zero for all sample sizes, thus confirming that in this setting the estimators are numerically identical as theory dictates. In the presence of censoring (whether right or interval censoring), the difference between the estimators decreased as sample size increased. Though not displayed in the figure, this trend continues and the proportion difference converges to zero with increasing sample size. Figure 1b shows that both the product and difference methods produced consistent estimators for the indirect effect. \\

\noindent Figure 2a shows the absolute proportion difference between the product and difference method under a Weibull model. In the absence of censoring, the proportion difference decreased as sample size increased but was still relatively large for small sample sizes. In the presence of right censoring, the proportion difference between the estimators was very large and did not decrease with increasing sample size. Figure 2b gives a summary of the proportion bias incurred by each estimator under the two censoring scenarios. For no censoring, both product and difference methods produced consistent estimators of the indirect effect. Under right censoring, the product method also produced an consistent indirect effect estimator across all sample sizes. In contrast, the difference method under right censoring had a proportion bias of about $45\%$, which does not appear to decrease with increasing sample size. The results from Figure 2b reveal that under right censoring, the difference method failed to be consistent for the indirect effect, with significant bias regardless of sample size. When there was no censoring, the difference method produced a consistent estimator of the indirect effect, although, in small samples the difference between the estimators was substantial. \\

\noindent In the above simulations, we only considered the indirect effect, but the results can be easily expanded to the total effect. For the normal model scenario, the total effect estimator is consistent. Thus, in both the absence and presence of censoring, the total effect can be estimated from the reduced form model ($\tau_a$ estimator) or by summation of direct and indirect effect estimators based on the product or difference method and the direct effect ($\beta_a$ estimator). For the Weibull model scenario, in the absence of censoring, the total effect can be estimated using either method, similar to the normal model scenario. In the presence of censoring, the total effect should only be estimated by the summation of the product method indirect effect estimator and the direct effect as the estimate of $\tau_a$ will be biased. \\ 

\noindent Figure A1 in the Appendix shows the Monte Carlo variances for each estimator discussed above. As expected, the variance of all estimators decreased towards zero as sample size increased.

\section*{Applications}

We considered a data application which combined two cohort studies of HIV-exposed persons: PHACS and PACTG 219C. The studies both followed perinatally HIV-exposed males and females upon entry into study and measured various outcomes. The outcome $T$ evaluated was age at sexual maturity for males only, which was subject to both interval and right censoring. Sexual maturity is defined as having reached stage 5 of the Tanner stage criteria for genitalia.$^{22}$ Previous research has modeled the outcome with a normal distribution, since age at attainment of pubertal milestones generally follows a normal distribution.$^{23,24}$ Thus, $T$ is adequately modeled as a normal outcome, and, therefore, we expect results for the normal model to apply. Of the 1,380 males in the sample, 28\% reached sexual maturity during follow-up and were subject to interval censoring; the remaining 72\% were right-censored. The exposure $A$ was binary perinatal HIV infection. The mediator $M$ was height age- and sex-adjusted Z-score (HTZ) at first visit occurring at age seven or older. \\

We adjusted for confounding by birth year and race. These correspond to $Z$ in models below. We fit a normal AFT model as age at sexual maturity is known to follow an approximately normal distribution. We used R to fit the following models in order to estimate the direct, indirect, and total effects using both the product and difference method. Note that R allows a ``Gaussian" distributed outcome in the survreg$^{12}$ function, so that the model can be written in terms of $T$ rather than $\log T$, though the same model can be obtained by fitting $\log T$ as a log-normal model for $\exp($age): \\

 			\begin{equation}
			T = \theta_0 + \theta_1  HIV  + \theta_2 HTZ + \theta_3^T Z + \varepsilon \label{normal1}
			\end{equation}
			\begin{equation}
			T = \beta_0 +  \beta_1  HIV + \beta_3^T Z + \xi \label{normal2}
			\end{equation}
			\begin{equation}
			HTZ = \alpha_0 + \alpha_1 HIV + \alpha_3^T Z +  \zeta \label{normal3}
			\end{equation}  
\noindent where $\varepsilon$, $\xi$, and $\zeta$ are all normally distributed variables with mean zero and unknown variance. 

	\footnotesize
            	\begin{table}[]
            	\centering
            	\caption{Normal AFT mediation model effect estimates for age at sexual maturity by perinatal HIV status ($n=1380$)}
            	\begin{tabular}{lccc}
		\hline
            	{\bf }                       & {\bf Estimate} & {\bf Standard Error} & {\bf 95\% CI}  \\ \hline
            	Direct                 & 4.14           & 3.55                 & (-2.82, 11.10)           \\
            	Indirect  (difference) & 2.90     & 0.97     & (1.00, 4.80)              \\ 
            	Indirect  (product)    & 2.99        & 0.65                 & (1.73, 4.26) \\ 
            	Total     (difference)             & 7.04           & 3.63           & (-0.07, 14.16)           \\ 
	         Total     (product)             &   7.13         &  3.28       & (0.70, 13.56)         \\ \hline
            	\end{tabular}
            	\end{table}
            	\normalsize

\vspace{.5cm} 
\noindent Table 1 displays effect estimates, their standard errors, and 95\% confidence intervals. Bootstrap estimates of standard errors were used for indirect (difference) and total (product) effect estimates. Our analysis indicated that HIV-infected youth had a 7.1 month delay in age at sexual maturity compared to uninfected youth; height Z-score accounting for approximately 40\% of the effect. There was a 3\% difference between the product and difference method estimators of the indirect effect. As discussed above, we do not expect numerical equivalence in the presence of censoring, though asymptotically both estimators should be consistent for the indirect effect. In addition, as we saw in the simulations, this sample size yielded a similar percent difference between estimators.  \\

\noindent In a second application, we combined 4 different randomized studies HIV-infected adults from the US-based AIDS Clinical Trials Group studies.$^{19-21}$ The binary exposure $A$ was treatment assignment at baseline to combination antiretroviral therapy versus monotherapy. The outcome $T$ was time to opportunistic infection or death and modeled as Weibull distributed. Out of 719 HIV-infected patients, 18\% experienced the outcome, and the remaining 82\% were right-censored. The mediator $M$ was change in viral load (log base 10 scale), which was measured at 8-weeks of follow-up. We excluded 12 people who had the event or were lost to follow-up within the first 8 weeks after treatment initiation and any subjects with missing values for change in viral load. Of the four studies, two randomized participants to either combination antiretroviral therapy versus monotherapy, while the other two studies randomized participants to two different types of monotherapy. As our comparison is no longer based on randomization, we adjusted for potential confounding by sex, weight, and IV drug use at baseline; these correspond to variable $Z$ in models below. \\

			\begin{equation}
			\log(T) = \theta_0 + \theta_1 Combination  + \theta_2 VL_{change} + \theta_3^T Z + \tilde{\varepsilon} \label{weibull1}
			\end{equation}
			\begin{equation}
			\log(T) = \beta_0 +  \beta_1  Combination + \beta_3^T Z +  \xi \label{weibull2}
			\end{equation}
			\begin{equation}
			VL_{change} = \alpha_0 + \alpha_1 Combination + \alpha_3^T Z + \tilde{\zeta} \label{weibull3}
			\end{equation} 

\noindent where $\tilde{\zeta}$ is normally distributed with mean zero and unknown variance and $\tilde{\varepsilon}$ and $\xi$ follow an extreme value distribution with unknown scale parameters.

\begin{table}[]
\centering
\caption{Weibull AFT mediation model effect estimates for time to death or OI by combination ARV treatment ($n=707$)}
\label{my-label}
\begin{tabular}{lccc}
\hline
{\bf }                       & {\bf Estimate (log-scale)} & {\bf Standard Error} & {\bf 95\% CI}  \\ \hline
Direct                 & 0.48          & 0.18                 & (0.13, 0.83)           \\
Indirect  (difference) & 0.22     &  0.04   &  (0.14, 0.29)         \\ 
Indirect  (product)    & 0.19        & 0.06                 & (0.08, 0.31) \\ 
Total (difference)            & 0.70           & 0.19                 &       (0.33, 1.07)    \\ 
Total (product)                & 0.67           &  0.18  &   (0.33, 1.02)         \\ \hline
\end{tabular}
\end{table}

\vspace{.5cm} 
\noindent  Table 2 displays effect estimates, their standard errors, and 95\% confidence intervals. Bootstrap estimates of standard errors were used for the indirect (difference) and total (product) effect estimates. Our estimates indicated a 2-fold increase in mean time to death or OI for adults starting combination ARVs as compared to monotherapy, but 28\% of this effect was mediated by decrease in viral load. The proportion difference between the two estimators was 12\%. As previously discussed, we do not expect exact numerical equivalence. Even asymptotically, we expect the estimators to be different, with only the product method yielding a consistent estimator for the indirect effect. As shown in the appendix, the total effect estimator via the reduced form AFT (``difference" in Table 2) will be biased, and the total effect (``product" in Table 2) is the only valid estimate for the total effect. Unlike the normal model discussed previously, we had no prior knowledge about the distribution of the time-to-event outcome. In order to assess goodness-of-fit, we compared the AIC of our full model with other potential distributions for the outcome: exponential, log-normal, log-logistic, and Rayleigh. We found that the Weibull time-to-event outcome provided the best fit as it had the lowest AIC. Furthermore, the Cox-Snell residual plot for our Weibull model showed that the fit was adequate as the residuals were relatively linear through the origin. \\ 

\section*{Conclusion}
In recent years, there has been an explosion of work to identify direct and indirect effects through causal mediation analysis in a variety of settings. The ease of estimating and interpreting direct and indirect effects from AFT models holds tremendous appeal to researchers, however, there are currently no explicit guidelines regarding possible complications due to censoring or truncation, two common phenomena in survival analysis. This paper offers such guidance, based on theoretical considerations, simulation studies, and two applications, establishing that the well-known equivalence of the product and difference approaches for estimating an indirect effect in linear models does not generally apply in the presence of censoring or truncation. \\

Specifically, we have formally established that the reduced form AFT model upon marginalizing over the mediator is mis-specified when the error distribution of the AFT model is not collapsible with respect of the error distribution of the mediator. In the presence of censoring or truncation, this misspecification can cause bias of the reduced form estimator of total effect, and therefore bias of the difference estimator of indirect effect. In the absence of censoring or truncation, the difference method yields a consistent estimator of the indirect effect.  However, the model-based variance of the difference methods is generally incorrect, since the information matrix is derived from an incorrect likelihood. In theory, one could correct this by using the nonparametric bootstrap or the sandwich variance estimator for inference. \\

The normal mediator-normal outcome model is an exception to the above phenomenon because the reduced form accelerated failure time model is correctly specified; thus, the product and difference method are both consistent for the indirect effect whether or not censoring or truncation is present. Crucially, consistency relies on both the mediator and the outcome following a normal distribution. If the mediator is not normally distributed, then the reduced form accelerated failure time model will be mis-specified. However, in the absence of censoring or truncation, we have shown that this form of model misspecification does not compromise consistency of the estimator of the indirect effect with either the product or difference method. Thus, the normality assumption of the mediator is only needed in the presence of censoring and truncation. \\

The normal mediator-normal outcome simulation study confirmed these results as the product and difference methods yielded consistent estimators of the indirect effect regardless of censoring. In addition, the Weibull simulation study confirmed that the difference method indirect effect estimator was biased and, thus, inconsistent in the presence of censoring, but consistent when there was no censoring. As shown in the appendix and our simulation results, under certain assumptions, the product method will always yield a consistent estimator of the indirect effect. Thus, we caution users against employing the difference method for AFT models, and generally recommend using the product method as it yields a consistent estimator of the indirect effect in any of the above scenarios. In addition, one could also use alternative semi-parametric methods that are less susceptible to modeling bias.$^{6,7,25}$ Regardless of the approach used, a careful evaluation of the distributional choice for models and assessment of potential confounders should always be conducted.

\newpage

\section*{Appendix: Mediation Analysis for Censored Survival Data under an Accelerated Failure Time Model}

\noindent For exposure $A$, mediator $M$ and outcome $T$, let $M(a)$ and
$T(a)=T(a,M(a))$ define the counterfactual mediator and outcome had exposure
taken value $a.$ Likewise, let $T(a,m)$ define the counterfactual outcome had
exposure and mediator taken the value $a$ and $m,$ respectively$.$ Finally let
$T(a,M(a^{\ast}))$ denote the counterfactual outcome had exposure taken value
$a$ and the mediator taken the value it would have under treatment $a^{\ast}.$
The average pure or natural direct effect on the log-additive scale is then
defined for $a\neq a^{\ast}:$

\[
NDE\left(  a,a^{\ast}\right)  =E\left\{ \log  T(a,M(a^{\ast}))\right\}  - E\left\{\log
T(a^{\ast})\right\}
\]
and the natural indirect effect is defined as
\[
NIE\left(  a,a^{\ast}\right)  =E\left\{  \log T(a)\right\}  -E\left\{\log 
T(a,M(a^{\ast}))\right\}
\]
Equivalently, we could write the above expressions conditioning on a set of confounders, $Z$. Throughout, we make the assumption:
\begin{equation} \tag{A1}
A\perp\!\!\!\perp\left\{  T(a,m),M(a)\right\} \mid Z  \label{randomize}%
\end{equation}
and \noindent we further suppose that we also have for all $a,a^{\ast}$ :%
\begin{equation}  \tag{A2}
T(a,m)\perp\!\!\!\perp M(a^{\ast})|A=a, Z \label{cross-world}%
\end{equation}
Under these assumptions, it follows that $NDE\left(  a,a^{\ast}\right)  $ and
$NIE\left(  a,a^{\ast}\right)  $ are identified empirically with$^2$
\[
E\left\{  \log T(a,M(a^{\ast})) \right\}  =\sum_{m,z}E\left\{ \log T|a,m,z \right\}
f\left(  m|a^{\ast},z\right) f\left( z\right)
\]
\noindent\textbf{Derivation of the indirect effect under an
	AFT\ model:} Suppose that the following accelerated failure time model holds,
\begin{equation}  \tag{A3}
\log T=\beta_{0}+\beta_{a}A+\beta_{m}M+\beta_z^T Z + \sigma\varepsilon
\label{conditional model}%
\end{equation}
where $\varepsilon$ is an independent residual of arbitrary distribution and not necessarily mean zero. \\

\noindent Assume that $M$ follows
\begin{equation} \tag{A4}
M=\alpha_{0}+\alpha_{a}A+\alpha_z^T Z+\xi
\label{Mmodel}%
\end{equation}
where $\xi$ is a mean zero error independent of $A$ and $Z$. \ Then,
\begin{align*}
E\left\{  \log T(a,M(a^{\ast})) \right\} & = \sum_{m,z}E\left\{   \log T|a,m,z\right\}  f\left(  m|a^{\ast},z\right)  \\
&  =\beta_{0}+ \beta_a a + \beta_m E(M \mid a^*,z) + \beta_z^T z + \sigma \epsilon \\
&  =\beta_{0}+ \beta_a a + \beta_m \alpha_0 + \beta_m \alpha_a a^* + \alpha_z^T z + \beta_z^T z + \sigma \epsilon
\end{align*}
which gives the following result,
\begin{align*}
NDE\left(  a,a^{\ast}\right)   &  =E\left\{  \log T(a,M(a^{\ast}))\right\}
- E\left\{  \log T(a^{\ast},M(a^{\ast}))\right\}  \\
&  =E\left\{  \log T(a,M(a^{\ast})) \mid Z \right\}
- E\left\{  \log T(a^{\ast},M(a^{\ast})) \mid Z \right\}  \\
&  =\beta_{a}\left(  a-a^{\ast}\right)
\end{align*}%
\begin{align*}
NIE\left(  a,a^{\ast}\right)   &  =E\left\{  \log T(a,M(a))\right\}  -
E\left\{  \log T(a,M(a^{\ast}))\right\}  \\
&  =E\left\{  \log T(a,M(a)) \mid Z \right\}  - E\left\{  \log T(a,M(a^{\ast})) \mid Z \right\}  \\
&  =\beta_{m}\alpha_{a}\left(  a-a^{\ast}\right)
\end{align*}

\noindent Note that under the AFT model one has the stronger result that at the individual level,

\begin{align*}
NDE\left(  a,a^{\ast}\right)   &  =  \log T(a,M(a^{\ast}))
-  \log T(a^{\ast},M(a^{\ast}))  \\
&  =\beta_{a}\left(  a-a^{\ast}\right)
\end{align*}
\begin{align*}
NIE\left(  a,a^{\ast}\right)   &  =\log T(a,M(a))  -
\log T(a,M(a^{\ast})) \\
&  =\beta_{m}\alpha_{a}\left(  a-a^{\ast}\right)
\end{align*}

\noindent For binary $A$ with $a=1$ and $a^*=0$, the indirect effect product method estimand is $\beta_{m}\alpha_{a}$ and the natural direct effect is $\beta_a$. The expression for the difference method is obtained from (\ref{conditional model}) and (\ref{Mmodel}): \\

\begin{equation} \tag{A5}
\begin{split}
\log T &  = \beta_{0} + \beta_{a}A + \beta_{m}M+ \beta_z^T Z + \sigma\varepsilon\\
& = \beta_{0} + \beta_{a}A + \beta_{m} (\alpha_{0}+\alpha_{a}A + \alpha_z^T Z +\xi) +\sigma\varepsilon\\
& = \beta_0 + \beta_{m} \alpha_{0} + (\beta_{a} + \beta_{m} \alpha_{a})A + (\beta_z^T + \alpha_z^T)Z + (\sigma\varepsilon + \beta_{m}\xi)\\
& =\beta_{0}^{\ast}+\tau_{a}A+ \beta_z^{\ast T}Z + \widetilde{\varepsilon}  \label{mis-model-1} 
\end{split}
\end{equation}

\noindent where $\widetilde{\varepsilon}$ follows the distribution given by the convolution of the density of $\sigma \varepsilon$ with that of $\beta_{m}\xi$,  which is independent of $A$ and $Z$. The total effect is given by $\tau_a$ and the indirect effect from the difference method is: 

\begin{equation} \tag{A6}
\tau_a - \beta_a = \alpha_a \beta_m  \label{A difference rule} 
\end{equation}

\noindent The difference method estimand is obtained by positing a second accelerated failure time model for $T$ as a function of $A$ and $Z$ only, which shall be referred to as the reduced form model and would typically be specified as followed:

\begin{equation} \tag{A7}
\log T=\beta_{0}^* + \tau_a A+ \beta_z^{\ast T}Z + \sigma \nu \label{A reduced model}
\end{equation}

\noindent where $\sigma$ is some unknown scale parameter to be estimated. Therefore, when using the difference method, one must specify the correct distribution of $\nu$  hoping to match that of $\widetilde{\varepsilon}$ in (\ref{mis-model-1}) -- failure to do so will result in model mis-specification. \\

\noindent\textbf{Evaluating consistency of the maximum likelihood estimator for $\tau_a$ under model mis-specification and right censoring:} Suppose that one mis-specifies the reduced form density of $T$ given $A$ and $Z$ from model (\ref{A reduced model}) with the density $f_{T}\left(t \mid X ; \alpha, \beta, \sigma \right) = f_{T}\left(t \mid X\right)$ and survival function $S_{T}\left(t \mid X ; \alpha, \beta, \sigma \right) = S_{T} \left( t \mid X \right)$. Let $X = (A, Z^T)^T, \beta = (\tau_a, \beta^{*T}_Z)$, and $\alpha$ is the intercept ($\beta^*_0$ above). We show below that the maximum likelihood estimator of $\beta$, and thus $\tau_a$, will be consistent in the absence of censoring. However, in the presence of censoring, the maximum likelihood estimator will not be consistent. We sketch the proof for the case of right censoring only. \\ 

\noindent The observed data is $\min (T,C)$ and $I(T\leq C)$ where $T$ is event time and $C$ is independent censoring time. The log likelihood for a single observation is: 

$$ \log \ell= I(T \leq C)\log f_{T}\left( T \mid X\right) +I(T>C)\log S_{T}\left(C \mid X \right)  $$

\noindent We can re-express this in terms of the rescaled residual error term, $T_0 = Te^{-\alpha -\beta X} = \exp(\sigma \varepsilon)$, which has density $f_0(T_0 \mid X) = f_0(T_0)$ because the residual error is independent of $X$, 

$$ f_T(t \mid X) = f_0(te^{-\alpha - \beta X}) e^{-\alpha - \beta X} $$

\noindent We can re-express the log likelihood: 

\begin{align*} 
\log \ell & = I(T \leq C)\log [  f_{0} (te^{-\alpha -\beta X}) e^{-\alpha -\beta X} ]+I(T>C)\log S_{0} (Ce^{ -\alpha - \beta X}) \\
& = I(T \leq C)\log (f_{0} (te^{\alpha -\beta X}))  - (\alpha + \beta X)  I(T \leq C) +I(T>C)\log (S_{0} (Ce^{- \beta X}))
\end{align*}

\noindent The score function of $\beta$, can be expressed as: 
\begin{align*} 
U_{\beta}(\beta, \alpha) & = \frac{d}{d \beta} \bigg[  I(T \leq C)\log (f_{0} (te^{\alpha -\beta X}))  - (\alpha + \beta X)  I(T \leq C) +I(T>C)\log( S_{0} (Ce^{- \beta X})) \bigg] \\
& = - X I(T \leq C) \frac{ \dot{f}_{0} (te^{-\alpha -\beta X}) te^{-\alpha -\beta X}}{ f_{0} (te^{-\alpha -\beta X}) }  - X  I(T \leq C)  +I(T>C) \frac{ \frac{d}{d\beta} S_{0} (Ce^{-\alpha - \beta X})}{S_0(Ce^{-\alpha -\beta X})}  \\
& = - X I(T \leq C) \frac{ \dot{f}_{0} (te^{-\alpha -\beta X}) te^{-\alpha -\beta X}}{ f_{0} (te^{-\alpha -\beta X}) }  - X  I(T \leq C)  \\
& \hspace{1cm} + \frac{ I(T>C) }{S_0(Ce^{-\alpha -\beta X})} \frac{d}{d\beta} \bigg(1 - \int_0^C f_0(te^{-\alpha -\beta X}) e^{-\alpha -\beta X} dt \bigg)  \\
& = - X I(T \leq C) \frac{ \dot{f}_{0} (te^{-\alpha -\beta X}) te^{-\alpha -\beta X}}{ f_{0} (te^{-\alpha -\beta X}) }  - X  I(T \leq C)  \\
& \hspace{1cm} + \frac{ I(T>C) }{S_0(Ce^{-\alpha -\beta X})}  \bigg(- \int_0^C \frac{\frac{d}{d\beta} [f_0(te^{-\alpha -\beta X}) e^{-\alpha -\beta X}]}{f_0(te^{-\alpha -\beta X})} f_0(te^{-\alpha -\beta X}) dt \bigg)  \\
& = -XI(T \leq C) \frac{ \dot{f}_{0} (te^{-\alpha -\beta X}) te^{-\alpha -\beta X}}{ f_{0} (te^{-\alpha -\beta X}) }  - X  I(T \leq C)  \\
& \hspace{1cm} + \frac{ I(T>C) }{S_0(Ce^{-\alpha -\beta X})}  \bigg(X \int_0^C \frac{ \dot{f}_0(te^{-\alpha -\beta X}) te^{-\alpha -\beta X} e^{-\alpha -\beta X} + f_0(e^{-\alpha -\beta X } )e^{-\alpha -\beta X}}{f_0(te^{-\alpha -\beta X})} f_0(te^{-\alpha - \beta X}) dt \bigg)  \\
\end{align*}

\noindent where $\dot{f}_{0}( \cdot )$ is the derivative of $f_0( \cdot )$ with respect to its argument. \\

\noindent Let ($\bar{\beta}, \bar{\alpha}$) denote the limiting value of the MLE, i.e. $(\hat{\alpha}, \hat{\beta}) \stackrel{P}{\to} (\bar{\beta}, \bar{\alpha})$ where $(\hat{\alpha}, \hat{\beta}) $ is the MLE. Then,

\[
E \bigg(
\begin{bmatrix}
U_{\beta} (\bar{\beta}, \bar{\alpha}) \\
U_{\alpha} (\bar{\beta}, \bar{\alpha})
\end{bmatrix}
\bigg)
= 0
\]

\noindent Now, we can take the expectation of the score of $\beta$ conditional on $C$ and $X$. Note that $f_0^*( \cdot )$ indicates the true law: 

\footnotesize
\begin{align*}
E[ U_{\beta}(\bar{\beta}, \bar{\alpha}) \mid C, X] & = E \bigg[ -XI(T \leq C) \frac{ \dot{f}_{0} (te^{-\bar{\alpha} - \bar{\beta} X}) te^{- \bar{\alpha} -\bar{\beta} X}}{ f_{0} (te^{- \bar{\alpha} - \bar{\beta} X}) }  - X  I(T \leq C)  \\
& \hspace{1cm} + \frac{ I(T>C) }{S_0(Ce^{-\bar{\alpha} -\bar{\beta} X})}  \bigg(X \int_0^C \frac{ \dot{f}_0(te^{- \bar{\alpha} - \bar{\beta} X}) te^{- \bar{\alpha} - \bar{\beta} X} e^{- \bar{\alpha} - \bar{\beta} X} + f_0(e^{- \bar{\alpha} -\bar{\beta} X } )e^{-\bar{\alpha} - \bar{\beta} X}}{f_0(te^{- \bar{\alpha} - \bar{\beta} X})} f_0(te^{- \bar{\alpha} - \bar{\beta} X}) dt \bigg) \mid C, X \bigg] \\
& =  \int_0^C (-X) \frac{ \dot{f}_{0}(te^{-\bar{\alpha} -\bar{\beta} X}) te^{-\bar{\alpha} -\bar{\beta} X}}{f_0(te^{-\bar{\alpha} - \bar{\beta} X})} f^*_0(te^{- \alpha -\beta X}) e^{-\alpha -\beta X} dt + \int_0^C (-X) f^*_0(te^{-\alpha -\beta X}) e^{-\alpha -\beta X} dt \\
& + \frac{ \int_C^{\infty} f^*_0(te^{-\alpha -\beta X}) e^{-\alpha -\beta X} dt }{S_0(Ce^{-\bar{\alpha} - \bar{\beta} X})}  \bigg(X \int_0^C \frac{ \dot{f}_0(te^{-\bar{\alpha} - \bar{\beta} X}) te^{-\bar{\alpha} -\bar{\beta} X} e^{-\bar{\alpha} -\beta X} + f_0(e^{-\bar{\alpha} - \bar{\beta} X } )e^{-\bar{\alpha} -\beta X}}{f_0(te^{-\bar{\alpha} -\bar{\beta} X})} f_0(te^{-\bar{\alpha} - \bar{\beta} X}) dt \bigg) \\
& = \int_0^C (-X) \frac{ \dot{f}_{0}(te^{-\bar{\alpha} -\bar{\beta} X}) te^{-\bar{\alpha} -\bar{\beta} X}}{f_0(te^{-\bar{\alpha} -\bar{\beta} X})} f^*_0(te^{-\alpha -\beta X}) e^{-\alpha -\beta X} dt + \int_0^C (-X) f^*_0(te^{-\alpha -\beta X}) e^{-\alpha -\beta X} dt \\
& \hspace{1cm} + \frac{ S^*_0(Ce^{-\alpha -\beta X})}{S_0(Ce^{-\bar{\alpha} - \bar{\beta X}})}  \bigg(X \int_0^C \frac{ \dot{f}_0(te^{-\bar{\alpha} -\bar{\beta} X}) te^{-\bar{\alpha} - \bar{\beta} X}  + f_0(e^{- \bar{\alpha} - \bar{\beta} X } )}{f_0(te^{-\bar{\alpha} -\bar{\beta} X})} f_0(te^{-\bar{\alpha} - \bar{\beta} X}) e^{-\bar{\alpha} -\bar{\beta} X} dt \bigg) \\
& = \int_0^C (-X) \frac{ \dot{f}_{0}(te^{-\bar{\alpha} -\bar{\beta} X}) te^{-\bar{\alpha} -\bar{\beta} X} + f_0(te^{-\bar{\alpha} -\bar{\beta} X})}{f_0(te^{-\bar{\alpha} -\bar{\beta} X})} f^*_0(te^{-\alpha -\beta X}) e^{-\alpha -\beta X} dt  \\
& \hspace{1cm} + \frac{ S^*_0(Ce^{-\alpha -\beta X})}{S_0(Ce^{-\bar{\alpha} -\bar{\beta }X})}  \bigg(X \int_0^C \frac{ \dot{f}_0(te^{-\bar{\alpha} -\bar{\beta} X}) te^{-\bar{\alpha} - \bar{\beta} X} + f_0(e^{-\bar{\alpha} -\bar{\beta} X } ) }{f_0(te^{-\bar{\alpha} -\bar{\beta} X})} f_0(te^{-\bar{\alpha} - \bar{\beta} X}) e^{-\bar{\alpha} -\bar{\beta} X} dt \bigg) \\
\end{align*}
\normalsize

\noindent Note that the conditional mean for the score of $\alpha $ is of similar form:

\begin{align*} E[U_{\alpha}(\bar{\beta},\bar{\alpha}) \mid C, X] & = \int_0^C - \frac{ \dot{f}_{0}(te^{-\bar{\alpha} -\bar{\beta} X}) te^{-\bar{\alpha} -\bar{\beta} X} + f_0(te^{-\bar{\alpha} -\bar{\beta} X})}{f_0(te^{-\bar{\alpha} -\bar{\beta} X})} f^*_0(te^{-\alpha -\beta X}) e^{-\alpha -\beta X} dt  \\
& \hspace{1cm} + \frac{ S^*_0(Ce^{-\alpha -\beta X})}{S_0(Ce^{-\bar{\alpha} -\bar{\beta }X})}  \bigg( \int_0^C \frac{ \dot{f}_0(te^{-\bar{\alpha} -\bar{\beta} X}) te^{-\bar{\alpha} - \bar{\beta} X} + f_0(e^{-\bar{\alpha} -\bar{\beta} X } ) }{f_0(te^{-\bar{\alpha} -\bar{\beta} X})} f_0(te^{-\bar{\alpha} - \bar{\beta} X}) e^{-\bar{\alpha} -\bar{\beta} X} dt \bigg)  \\
\end{align*}
\noindent Let $p$ be the mean vector for the vector $X$. Noting that $E \big[ U_{\beta}(\bar{\beta},\bar{\alpha})] = 0$ and $E \big[ U_{\alpha}(\bar{\beta},\bar{\alpha})] = 0$, we can write, 
\begin{align*}
E \big[ U_{\beta}(\bar{\beta},\bar{\alpha})] & = E \big[ U_{\beta}(\bar{\beta},\bar{\alpha}) - p U_{\alpha}(\bar{\beta},\bar{\alpha}) \big] \\
& =  E \bigg[ - (X-p) \int_0^C \frac{ \dot{f}_{0}(te^{-\bar{\alpha} -\bar{\beta} X}) te^{-\bar{\alpha} -\bar{\beta} X} + f_0(te^{-\bar{\alpha} -\bar{\beta} X})}{f_0(te^{-\bar{\alpha} -\bar{\beta} X})} f^*_0(te^{-\alpha -\beta X}) e^{-\alpha -\beta X} dt  \\
& \hspace{.5cm} + (X-p)\frac{ S^*_0(Ce^{-\alpha -\beta X})}{S_0(Ce^{-\bar{\alpha} -\bar{\beta }X})}  \bigg( \int_0^C \frac{ \dot{f}_0(te^{-\bar{\alpha} -\bar{\beta} X}) te^{-\bar{\alpha} - \bar{\beta} X} + f_0(e^{-\bar{\alpha} -\bar{\beta} X } ) }{f_0(te^{-\bar{\alpha} -\bar{\beta} X})} f_0(te^{-\bar{\alpha} - \bar{\beta} X}) e^{-\bar{\alpha} -\bar{\beta} X} dt \bigg) \bigg] \\
\end{align*}
\noindent We will now plug in the true values to assess whether we get an unbiased score equation under model mis-specification for $\beta$, i.e. $E \big[ U_{\beta}(\beta,\bar{\alpha})] = 0$. Suppose that $\bar{\beta} = \beta$ : 
\begin{align*}
& =  E \bigg[ - (X-p) \int_0^C \frac{ \dot{f}_{0}(te^{-\bar{\alpha} -\beta X}) te^{-\bar{\alpha} - \beta X} + f_0(te^{-\bar{\alpha} -\beta X})}{f_0(te^{-\bar{\alpha} - \beta X})} f^*_0(te^{-\alpha -\beta X}) e^{-\alpha -\beta X} dt  \\
& \hspace{1cm} + (X-p)\frac{ S^*_0(Ce^{-\alpha -\beta X})}{S_0(Ce^{-\bar{\alpha} -\beta X})}  \bigg( \int_0^C \frac{ \dot{f}_0(te^{-\bar{\alpha} - \beta X}) te^{-\bar{\alpha} - \beta X} + f_0(e^{-\bar{\alpha} -\beta X } ) }{f_0(te^{-\bar{\alpha} -\beta X})} f_0(te^{-\bar{\alpha} - \beta X}) e^{- \bar{\alpha} - \beta X} dt \bigg)  \bigg] \\
& = E\bigg[ -(X-p) \int_0^{Ce^{-\beta X}}  \frac{ \dot{f}_{0}(e^{-\bar{\alpha}}u) e^{-\bar{\alpha}}u + f_0(e^{-\bar{\alpha}}u)}{f_0(e^{-\bar{\alpha}}u)} f^*_0(e^{-\alpha}u) e^{-\alpha} du   \\
& \hspace{1cm} + (X-p) \frac{ S^*_0(Ce^{-\alpha -\beta X})}{S_0(Ce^{-\bar{\alpha} -\beta X})} \int_0^{Ce^{-\beta X}} \frac{ \dot{f}_0(e^{-\bar{\alpha}}u) e^{-\bar{\alpha}}u + f_0(e^{-\bar{\alpha}}u)}{f_0(e^{-\bar{\alpha}}u)} f_0(e^{-\bar{\alpha}}u) e^{-\bar{\alpha}}du \bigg] \\
& = E \bigg[ (p-X) \int_0^{Ce^{-\beta X}} \big[ 1-\frac{ S^*_0(Ce^{-\alpha -\beta X}) f_0(e^{-\bar{\alpha}}u)e^{-\bar{\alpha}}}{S_0(Ce^{-\bar{\alpha} -\beta X}) f^*_0(e^{-\alpha}u)e^{-\alpha}}  \big] \frac{ \dot{f}_{0}(e^{-\bar{\alpha}}u) u + f_0(e^{-\bar{\alpha}}u)}{f_0(e^{-\bar{\alpha}}u)} f^*_0(e^{-\alpha}u)e^{-\alpha} du \bigg]  \\
& = \int_0^{\infty} E \bigg[ (p-X)  \big[1 - \frac{ S^*_0(Ce^{-\alpha -\beta X}) f_0(e^{-\bar{\alpha}}u)e^{-\bar{\alpha}}}{S_0(Ce^{-\bar{\alpha} -\beta X}) f^*_0(e^{-\alpha}u)e^{-\alpha}} \big] I(u < Ce^{ -\beta X}) \bigg] \frac{ \dot{f}_{0}(e^{-\bar{\alpha}}u) u + f_0(e^{-\bar{\alpha}}u)}{f_0(e^{-\bar{\alpha}}u)} f^*_0(e^{-\alpha}u)e^{-\alpha} du \hspace{.2cm} \textrm{(A8)} 
\end{align*}

\noindent If there is no right censoring ($C \to \infty$), and for every value of $x$:

\begin{align*} 
& I\left( u<Ce^{-\beta x} \right) \to 1 \\
& \frac{S_{T}^{\ast }\left( Ce^{-\alpha - \beta x}  \right) }{S_{T}\left(Ce^{-\bar{\alpha} - \beta x} \right) } \to 1 \\
\end{align*} 

\noindent in which case the expectation evaluates to zero, and the score for $\beta$ is unbiased. Additionally, if the model is not mis-specified, so that $S_0^*(\cdot) = S_0(\cdot)$ and $f_0^*(\cdot) = f_0(\cdot)$, then the score for $\beta$ will also be unbiased regardless of censoring. Thus, the association of $X$ with $T$ is consistent in the absence of censoring. However, in the presence of censoring, the above will not necessarily evaluate to zero. To show this, we consider a special case when $X$ is binary: 

\begin{align*}
& = \int_0^{\infty}  E \bigg[ (p-X) X \bigg( \big[1 - \frac{ S^*_0(Ce^{-\alpha -\beta }) f_0(e^{-\bar{\alpha}}u)e^{-\bar{\alpha}}}{S_0(Ce^{-\bar{\alpha} -\beta}) f^*_0(e^{-\alpha}u)e^{-\alpha}} \big] I(u < Ce^{ -\beta}) - \big[1 - \frac{ S^*_0(Ce^{-\alpha}) f_0(e^{-\bar{\alpha}}u)e^{-\bar{\alpha}}}{S_0(Ce^{-\bar{\alpha}}) f^*_0(e^{-\alpha}u)e^{-\alpha}} \big] I(u < C) \bigg) \\
& \hspace{1cm} + \big[1 - \frac{ S^*_0(Ce^{-\alpha}) f_0(e^{-\bar{\alpha}}u)e^{-\bar{\alpha}}}{S_0(Ce^{-\bar{\alpha}}) f^*_0(e^{-\alpha}u)e^{-\alpha}} \big] I(u < C) \bigg] \frac{ \dot{f}_{0}(e^{-\bar{\alpha}}u) u + f_0(e^{-\bar{\alpha}}u)}{f_0(e^{-\bar{\alpha}}u)} f^*_0(e^{-\alpha}u)e^{-\alpha} du  \\
& = \int_0^{\infty}  E \bigg[ (X-p) X \bigg( \frac{ S^*_0(Ce^{-\alpha -\beta }) }{S_0(Ce^{-\bar{\alpha} -\beta}) } I(u < Ce^{ -\beta}) -  \frac{ S^*_0(Ce^{-\alpha}) }{S_0(Ce^{-\bar{\alpha}}) }  I(u < C) \bigg) \frac{f_0(e^{-\bar{\alpha}}u)e^{-\bar{\alpha}}}{f^*_0(e^{-\alpha}u)e^{-\alpha}} \bigg] \\
& \hspace{1cm} \times \frac{ \dot{f}_{0}(e^{-\bar{\alpha}}u) u + f_0(e^{-\bar{\alpha}}u)}{f_0(e^{-\bar{\alpha}}u)} f^*_0(e^{-\alpha}u)e^{-\alpha} du  \\
& =  p(1-p) \int_0^{\infty}  E \bigg[  \frac{ S^*_0(Ce^{-\alpha -\beta }) }{S_0(Ce^{-\bar{\alpha} -\beta}) } I(u < Ce^{ -\beta}) -  \frac{ S^*_0(Ce^{-\alpha}) }{S_0(Ce^{-\bar{\alpha}}) }  I(u < C) \bigg] e^{-\bar{\alpha}}[\dot{f}_{0}(e^{-\bar{\alpha}}u)u + f_0(e^{-\bar{\alpha}} u )] du \\
\end{align*}

\noindent The above expression will generally be nonzero except at exceptional laws, such as when $\beta =0$. Therefore, in the presence of model mis-specification, censoring, and a non-null effect, the MLE of $\tau_a$ will not be consistent. \\

\noindent\textbf{Evaluating consistency of the maximum likelihood estimator for $\tau_a$ under model mis-specification and left truncation:} Suppose that one mis-specifies the reduced form density of $T$ given $A$ and $Z$ from model (\ref{A reduced model}) with the density $f_{T}\left(t \mid X ; \alpha, \beta, \sigma \right) = f_{T}\left(t \mid X\right)$ and survival function $S_{T}\left(t \mid X ; \alpha, \beta, \sigma \right) = S_{T} \left( t \mid X \right)$. Let $X = (A, Z^T)^T, \beta = (\tau_a, \beta^{*T}_Z)$, and $\alpha$ is the intercept ($\beta^*_0$ above). We show below that the maximum likelihood estimator of $\beta$, and thus $\tau_a$, will be consistent in the absence of left truncation. However, in the presence of left truncation, the maximum likelihood estimate will not be consistent.  \\ 

\noindent  Let $T$ be left truncated at $V$ such that we consider $T \mid T \geq V$ assuming that the truncation time is independent of $T$ and $X$, but otherwise follows an unrestricted density. The log likelihood for a single observation subject to left truncation is: 

$$ \log \ell= \log f_{T}\left( T \mid X\right) - \log S_{T}\left(V \mid X \right)  $$

\noindent We can re-express this in terms of the rescaled residual error term, $T_0 = Te^{-\alpha -\beta X} = \exp(\sigma \varepsilon)$, which has density $f_0(T_0 \mid X) = f_0(T_0)$ because the residual error term is independent of $X$, the following way:

$$ f_T(t \mid X) = f_0(te^{-\alpha - \beta X}) e^{-\alpha - \beta X} $$

\noindent We can re-express the log likelihood: 

\begin{align*} 
\log \ell & = \log [  f_{0} (te^{-\alpha -\beta X}) e^{-\alpha -\beta X} ] - \log (S_{0} (Ve^{ -\alpha - \beta X})) \\
& = \log [  f_{0} (te^{-\alpha -\beta X}) ] - (\alpha +\beta X) - \log (S_{0} (Ve^{ -\alpha - \beta X}))
\end{align*}

\noindent The score function of $\beta$ can be expressed as: 

\begin{align*} 
U_{\beta}(\alpha,\beta)& = - X \frac{ \dot{f}_{0} (te^{-\alpha -\beta X}) te^{-\alpha -\beta X}}{ f_{0} (te^{-\alpha -\beta X}) }  - X   - \frac{ \frac{d}{d\beta} S_{0} (Ve^{-\alpha - \beta X})}{S_0(V^{-\alpha -\beta X})}  \\
& = - X  \frac{ \dot{f}_{0} (te^{-\alpha -\beta X}) te^{-\alpha -\beta X}}{ f_{0} (te^{-\alpha -\beta X}) }  - X   - \frac{ 1 }{S_0(Ve^{-\alpha -\beta X})} \frac{d}{d\beta} \bigg(1 - \int_0^V f_0(te^{-\alpha -\beta X}) e^{-\alpha -\beta X} dt \bigg)  \\
& = - X  \frac{ \dot{f}_{0} (te^{-\alpha -\beta X}) te^{-\alpha -\beta X}}{ f_{0} (te^{-\alpha -\beta X}) }  - X   - \frac{ 1 }{S_0(Ve^{-\alpha -\beta X})} \frac{d}{d\beta} \bigg(\int_V^{\infty} f_0(te^{-\alpha -\beta X}) e^{-\alpha -\beta X} dt \bigg)  \\
& = - X  \frac{ \dot{f}_{0} (te^{-\alpha -\beta X}) te^{-\alpha -\beta X}}{ f_{0} (te^{-\alpha -\beta X}) }  - X   - \frac{ 1 }{S_0(Ve^{-\alpha -\beta X})}  \bigg(\int_V^{\infty} \frac{\frac{d}{d\beta} [f_0(te^{-\alpha -\beta X}) e^{-\alpha -\beta X}]}{f_0(te^{-\alpha -\beta X})} f_0(te^{-\alpha -\beta X}) dt \bigg)  \\
& = -X  \frac{ \dot{f}_{0} (te^{-\alpha -\beta X}) te^{-\alpha -\beta X}}{ f_{0} (te^{-\alpha -\beta X}) }  - X  \\
& \hspace{1.2cm} + \frac{ 1 }{S_0(Ve^{-\alpha -\beta X})}  \bigg(X \int_V^{\infty} \frac{ \dot{f}_0(te^{-\alpha -\beta X}) te^{-\alpha -\beta X} e^{-\alpha -\beta X} + f_0(e^{-\alpha -\beta X } )e^{-\alpha -\beta X}}{f_0(te^{-\alpha -\beta X})}f_0(te^{-\alpha -\beta X})  dt \bigg)  \\
\end{align*}

\noindent where $\dot{f}_{0}$ is the derivative of $f_0$ with respect to its argument.  \\

\noindent Let ($\bar{\beta}, \bar{\alpha}$) denote the limiting value of the MLE, i.e. $(\hat{\alpha}, \hat{\beta}) \stackrel{P}{\to} (\bar{\beta}, \bar{\alpha})$ where $(\hat{\alpha}, \hat{\beta}) $ is the MLE. Then,

\[
E \bigg(
\begin{bmatrix}
U_{\beta} (\bar{\beta}, \bar{\alpha}) \\
U_{\alpha} (\bar{\beta}, \bar{\alpha})
\end{bmatrix}
\bigg)
= 0
\]

\noindent Now, we can take the expectation of the score of $\beta$ conditional on $X$ with respect to the density of $T \mid T > V$. Note that $f_0^*( \cdot )$ indicates the true law: 

\footnotesize
\begin{align*}
E[ U_{\beta}(\bar{\beta},\bar{\alpha}) \mid X, V] & = \int_V^{\infty} (-X) \frac{ \dot{f}_{0}(te^{-\bar{\alpha} - \bar{\beta} X}) te^{- \bar{\alpha} - \bar{\beta} X}}{f_0(te^{- \bar{\alpha} -\bar{\beta} X})} \frac{f^*_0(te^{-\alpha -\beta X})}{S^*_0(Ve^{-\alpha - \beta X})} e^{-\alpha -\beta X} dt + \int_V^{\infty} (-X) \frac{f^*_0(te^{-\alpha -\beta X})}{S^*_0(Ve^{-\alpha - \beta X})} e^{-\alpha -\beta X} dt \\
& \hspace{1cm} + \frac{ \int_V^{\infty}\frac{f^*_0(te^{-\alpha -\beta X})}{S^*_0(Ve^{-\alpha - \beta X})} e^{-\alpha -\beta X} dt }{S_0(Ve^{-\bar{\alpha} -\bar{\beta} X})}  \bigg(X \int_V^{\infty} \frac{ \dot{f}_0(te^{-\bar{\alpha} -\bar{\beta} X}) te^{-\bar{\alpha} - \bar{\beta} X} e^{- \bar{\alpha} - \bar{\beta} X} + f_0(e^{-\bar{\alpha} -\bar{\beta} X } )e^{-\bar{\alpha} -\bar{\beta} X}}{f_0(te^{-\bar{\alpha} -\bar{\beta} X})} f_0(te^{-\bar{\alpha} -\bar{\beta} X}) dt \bigg) \\
& = \int_V^{\infty} (-X) \frac{ \dot{f}_{0}(te^{-\bar{\alpha} - \bar{\beta} X}) te^{- \bar{\alpha} - \bar{\beta} X}}{f_0(te^{-\bar{\alpha} - \bar{\beta} X})} \frac{f^*_0(te^{-\alpha -\beta X})}{S^*_0(Ve^{-\alpha - \beta X})} e^{-\alpha -\beta X} dt + \int_V^{\infty} (-X) \frac{f^*_0(te^{-\alpha -\beta X})}{S^*_0(Ve^{-\alpha - \beta X})} e^{-\alpha -\beta X} dt \\
& \hspace{1cm} + \frac{ 1}{S_0(Ve^{-\bar{\alpha} - \bar{\beta} X})} \bigg(X \int_V^{\infty} \frac{ \dot{f}_0(te^{-\bar{\alpha} - \bar{\beta} X}) te^{-\bar{\alpha} -\bar{\beta} X} e^{-\bar{\alpha} - \bar{\beta} X} + f_0(e^{-\bar{\alpha} -\bar{\beta} X } )e^{-\bar{\alpha} -\bar{\beta} X}}{f_0(te^{-\bar{\alpha} -\bar{\beta} X})} f_0(te^{-\bar{\alpha} -\bar{\beta} X}) dt \bigg)  \\
& = \frac{1}{S^*_0(Ve^{-\alpha - \beta X})}\int_V^{\infty} (-X) \frac{ \dot{f}_{0}(te^{-\bar{\alpha} - \bar{\beta} X}) te^{- \bar{\alpha} - \bar{\beta} X} + f_0(te^{- \bar{\alpha} - \bar{\beta} X})}{f_0(te^{- \bar{\alpha} - \bar{\beta} X})} f^*_0(te^{-\alpha -\beta X}) e^{-\alpha -\beta X} dt  \\
& \hspace{1cm} + \frac{ 1}{S_0(Ve^{-\bar{\alpha} - \bar{\beta} X})}  \bigg(X \int_V^{\infty} \frac{ \dot{f}_0(te^{- \bar{\alpha} - \bar{\beta} X}) te^{-\bar{\alpha} - \bar{\beta} X} + f_0(e^{- \bar{\alpha} - \bar{\beta} X } ) }{f_0(te^{- \bar{\alpha} - \bar{\beta} X})} f_0(te^{- \bar{\alpha} - \bar{\beta} X}) e^{-\bar{\alpha} - \bar{\beta} X} dt \bigg) \\
\end{align*}
\normalsize

\noindent Note that the conditional mean for the score of $\alpha$ is of similar form and satisfies:

\begin{align*}
E[ U_{\alpha}(\bar{\beta},\bar{\alpha}) \mid X, V] & = \frac{1}{S^*_0(Ve^{-\alpha - \beta X})}\int_V^{\infty} (-X) \frac{ \dot{f}_{0}(te^{-\bar{\alpha} - \bar{\beta} X}) te^{- \bar{\alpha} - \bar{\beta} X} + f_0(te^{- \bar{\alpha} - \bar{\beta} X})}{f_0(te^{- \bar{\alpha} - \bar{\beta} X})} f^*_0(te^{-\alpha -\beta X}) e^{-\alpha -\beta X} dt  \\
& \hspace{1cm} + \frac{ 1}{S_0(Ve^{-\bar{\alpha} - \bar{\beta} X})}  \bigg(X \int_V^{\infty} \frac{ \dot{f}_0(te^{- \bar{\alpha} - \bar{\beta} X}) te^{-\bar{\alpha} - \bar{\beta} X} + f_0(e^{- \bar{\alpha} - \bar{\beta} X } ) }{f_0(te^{- \bar{\alpha} - \bar{\beta} X})} f_0(te^{- \bar{\alpha} - \bar{\beta} X}) e^{-\bar{\alpha} - \bar{\beta} X} dt \bigg) \\
\end{align*} 

\noindent Let $p$ be the mean vector for the vector $X$. Noting that $E \big[ U_{\beta}(\bar{\beta},\bar{\alpha})] = 0$ and $E \big[ U_{\alpha}(\bar{\beta},\bar{\alpha})] = 0$, we can write, 
\begin{align*}
E \big[ U_{\beta}(\bar{\beta},\bar{\alpha})] & = E \big[ U_{\beta}(\bar{\beta},\bar{\alpha}) - p U_{\alpha}(\bar{\beta},\bar{\alpha}) \big] \\
& =  E \bigg[ - (X-p) \frac{1}{S^*_0(Ve^{-\alpha - \beta X})}\int_V^{\infty} \frac{ \dot{f}_{0}(te^{-\bar{\alpha} - \bar{\beta} X}) te^{- \bar{\alpha} - \bar{\beta} X} + f_0(te^{- \bar{\alpha} - \bar{\beta} X})}{f_0(te^{- \bar{\alpha} - \bar{\beta} X})} f^*_0(te^{-\alpha -\beta X}) e^{-\alpha -\beta X} dt  \\
& \hspace{.5cm} + (X-p)\frac{ 1}{S_0(Ve^{-\bar{\alpha} - \bar{\beta} X})}  \bigg( \int_V^{\infty} \frac{ \dot{f}_0(te^{- \bar{\alpha} - \bar{\beta} X}) te^{-\bar{\alpha} - \bar{\beta} X} + f_0(e^{- \bar{\alpha} - \bar{\beta} X } ) }{f_0(te^{- \bar{\alpha} - \bar{\beta} X})} f_0(te^{- \bar{\alpha} - \bar{\beta} X}) e^{-\bar{\alpha} - \bar{\beta} X} dt \bigg) \bigg] \\
\end{align*}
\noindent We will now plug in the true values to assess whether we get an unbiased score equation under model mis-specification for $\beta$, i.e. $E \big[ U_{\beta}(\beta,\bar{\alpha})] = 0$. Suppose that $\bar{\beta} = \beta$ :

\begin{align*}
& =  E \bigg[ - (X-p) \frac{1}{S^*_0(Ve^{-\alpha - \beta X})}\int_V^{\infty} \frac{ \dot{f}_{0}(te^{-\bar{\alpha} - \beta X}) te^{- \bar{\alpha} - \beta X} + f_0(te^{- \bar{\alpha} - \beta X})}{f_0(te^{- \bar{\alpha} - \beta X})} f^*_0(te^{-\alpha -\beta X}) e^{-\alpha -\beta X} dt  \\
& \hspace{1cm} + (X-p)\frac{ 1}{S_0(Ve^{-\bar{\alpha} - \beta X})}  \bigg( \int_V^{\infty} \frac{ \dot{f}_0(te^{- \bar{\alpha} - \beta X}) te^{-\bar{\alpha} - \beta X} + f_0(e^{- \bar{\alpha} - \beta X } ) }{f_0(te^{- \bar{\alpha} - \beta X})} f_0(te^{- \bar{\alpha} - \beta X}) e^{-\bar{\alpha} - \beta X} dt \bigg) \bigg] \\
& = E \bigg[ - (X-p) \frac{1}{S^*_0(Ve^{-\alpha - \beta X})}\int_{Ve^{-\beta X}}^{\infty}  \frac{ \dot{f}_{0}(e^{-\bar{\alpha}}u) e^{- \bar{\alpha}}u + f_0(e^{- \bar{\alpha}}u)}{f_0(e^{- \bar{\alpha}}u)} f^*_0(e^{-\alpha}u) e^{-\alpha} du  \\
& \hspace{1cm} + (X-p)\frac{ 1}{S_0(Ve^{-\bar{\alpha} - \beta X})}  \bigg( \int_{Ve^{-\beta X}}^{\infty} \frac{ \dot{f}_0(e^{- \bar{\alpha}}u) e^{-\bar{\alpha}}u + f_0(e^{- \bar{\alpha}}u ) }{f_0(e^{- \bar{\alpha}}u)} f_0(e^{- \bar{\alpha}}u) e^{-\bar{\alpha}} du \bigg) \bigg] \\
& = E \bigg[ - (X-p)\int_{Ve^{-\beta X}}^{\infty} \big[ \frac{1}{S^*_0(Ve^{-\alpha - \beta X})} -  \frac{ f_0(e^{-\bar{\alpha}}u) e^{-\bar{\alpha}} }{ f^*_0(e^{-\alpha}u) e^{-\alpha} S_0(Ve^{-\bar{\alpha} - \beta X})} \big] \\
& \hspace{8cm} \times \frac{ \dot{f}_{0}(e^{-\bar{\alpha}}u) e^{- \bar{\alpha}}u + f_0(e^{- \bar{\alpha}}u)}{f_0(e^{- \bar{\alpha}}u)} f^*_0(e^{-\alpha}u) e^{-\alpha} du \bigg] \\
& = \int_{0}^{\infty} E \bigg[ - (X-p) \big[ \frac{1}{S^*_0(Ve^{-\alpha - \beta X})} -  \frac{ f_0(e^{-\bar{\alpha}}u) e^{-\bar{\alpha}} }{ f^*_0(e^{-\alpha}u) e^{-\alpha} S_0(Ve^{-\bar{\alpha} - \beta X})} \big] I(u > Ve^{-\beta X} )\bigg] \\
& \hspace{8cm} \times \frac{ \dot{f}_{0}(e^{-\bar{\alpha}}u) e^{- \bar{\alpha}}u + f_0(e^{- \bar{\alpha}}u)}{f_0(e^{- \bar{\alpha}}u)} f^*_0(e^{-\alpha}u) e^{-\alpha} du  \hspace{1cm} \textrm{(A9)} \\
\end{align*}

\noindent If there is no left truncation ($V=0$), and for every value of $x$:

\begin{align*} 
& I ( u>Ve^{- \beta x} ) = I\left( u>0 \right)= 1 \\
&  \frac{1}{S^*_0(Ve^{-\alpha - \beta X})} -  \frac{ f_0(e^{-\bar{\alpha}}u) e^{-\bar{\alpha}} }{ f^*_0(e^{-\alpha}u) e^{-\alpha} S_0(Ve^{-\bar{\alpha} - \beta X})} = 1 - \frac{f_0(e^{-\bar{\alpha}}u) e^{-\bar{\alpha}}}{f^*_0(e^{-\alpha}u) e^{-\alpha} }\\
\end{align*} 

\noindent in which case the expectation evaluates to zero, and the score for $\beta$ is unbiased. Additionally, if the model is not mis-specified, so that $S_0^*(\cdot) = S_0(\cdot)$ and $f_0^*(\cdot) = f_0(\cdot)$, then the score for $\beta$ will also be unbiased. Thus, the association of $X$ with $T$ is consistent in the absence of censoring. However, in the presence of left truncation, the above will not necessarily evaluate to zero. To show this, we consider a special case when $X$ is binary: 

\begin{align*}
& = \int_{0}^{\infty} E \bigg[ - (X-p) X \bigg( \big[ \frac{1}{S^*_0(Ve^{-\alpha - \beta})} -  \frac{ f_0(e^{-\bar{\alpha}}u) e^{-\bar{\alpha}} }{ f^*_0(e^{-\alpha}u) e^{-\alpha} S_0(Ve^{-\bar{\alpha} - \beta})} \big] I(u > Ve^{-\beta} ) \\
& \hspace{2cm} -  \big[ \frac{1}{S^*_0(Ve^{-\alpha})} -  \frac{ f_0(e^{-\bar{\alpha}}u) e^{-\bar{\alpha}} }{ f^*_0(e^{-\alpha}u) e^{-\alpha} S_0(Ve^{-\bar{\alpha}})} \big] I(u > V ) \bigg) \\
& \hspace{3cm} + \frac{ f_0(e^{-\bar{\alpha}}u) e^{-\bar{\alpha}} }{ f^*_0(e^{-\alpha}u) e^{-\alpha} S_0(Ve^{-\bar{\alpha}})} \big] I(u > V ) \bigg]  \frac{ \dot{f}_{0}(e^{-\bar{\alpha}}u) e^{- \bar{\alpha}}u + f_0(e^{- \bar{\alpha}}u)}{f_0(e^{- \bar{\alpha}}u)} f^*_0(e^{-\alpha}u) e^{-\alpha} du  \\
& = \int_{0}^{\infty} E \bigg[ - (X-p) X \bigg( \big[ \frac{1}{S^*_0(Ve^{-\alpha - \beta})} -  \frac{ f_0(e^{-\bar{\alpha}}u) e^{-\bar{\alpha}} }{ f^*_0(e^{-\alpha}u) e^{-\alpha} S_0(Ve^{-\bar{\alpha} - \beta})} \big] I(u > Ve^{-\beta} ) \\
& \hspace{2cm} -  \big[ \frac{1}{S^*_0(Ve^{-\alpha})} -  \frac{ f_0(e^{-\bar{\alpha}}u) e^{-\bar{\alpha}} }{ f^*_0(e^{-\alpha}u) e^{-\alpha} S_0(Ve^{-\bar{\alpha}})} \big] I(u > V ) \bigg) \bigg]  \frac{ \dot{f}_{0}(e^{-\bar{\alpha}}u) e^{- \bar{\alpha}}u + f_0(e^{- \bar{\alpha}}u)}{f_0(e^{- \bar{\alpha}}u)} f^*_0(e^{-\alpha}u) e^{-\alpha} du  \\
& = -p(1-p) \int_{0}^{\infty} E \bigg[ \big[ \frac{1}{S^*_0(Ve^{-\alpha - \beta})} -  \frac{ f_0(e^{-\bar{\alpha}}u) e^{-\bar{\alpha}} }{ f^*_0(e^{-\alpha}u) e^{-\alpha} S_0(Ve^{-\bar{\alpha} - \beta})} \big] I(u > Ve^{-\beta} ) \\
& \hspace{2cm} - \big[ \frac{1}{S^*_0(Ve^{-\alpha})} -  \frac{ f_0(e^{-\bar{\alpha}}u) e^{-\bar{\alpha}} }{ f^*_0(e^{-\alpha}u) e^{-\alpha} S_0(Ve^{-\bar{\alpha}})} \big] I(u > V ) \bigg]  \frac{ \dot{f}_{0}(e^{-\bar{\alpha}}u) e^{- \bar{\alpha}}u + f_0(e^{- \bar{\alpha}}u)}{f_0(e^{- \bar{\alpha}}u)} f^*_0(e^{-\alpha}u) e^{-\alpha} du  \\
\end{align*}

\noindent The above expression will generally be nonzero except at exceptional laws, such as when $\beta = 0$. Therefore, in the presence of model mis-specification, censoring, and a non-null effect, the MLE of $\tau_a$ will not be consistent. \\

\noindent\textbf{An issue with equivalence of the product and difference method indirect effect under
	a AFT\ model with a Weibull outcome, no censoring:} Consider model (\ref{conditional model}) and (\ref{Mmodel}), where
$\varepsilon$ follows an extreme value distribution and $\xi$ is normally
distributed. Then the implied reduced form model is given by:
\begin{equation} \tag{A10}
\begin{split}
\log T &  = \beta_{0} + \beta_{a}A + \beta_{m}M+ \beta_z^T Z+ \sigma\varepsilon\\
& = \beta_{0} + \beta_{a}A + \beta_{m} (\alpha_{0}+\alpha_{a}A+\alpha_z^T Z \xi) + \beta_z^T Z +\sigma\varepsilon\\
& = \beta_0 + \beta_{m} \alpha_{0} + (\beta_{a} + \beta_{m} \alpha_{a})A + (\alpha_z^T + \beta_z^T)Z + (\sigma\varepsilon + \beta_{m}\xi)\\
& =\beta_{0}^{\ast}+\tau_{a}A + \beta_z^{\ast T}Z + \widetilde{\varepsilon}  \label{mis-model}
\end{split}
\end{equation}

\noindent where $\beta_{0}^{\ast}=\beta_{m}\alpha_{0}+\beta_{0},$ $\beta_z^{\ast T} = \alpha_z^T + \beta_z^T$, 
$\widetilde{\varepsilon}=  \beta_{m}\xi+\sigma\varepsilon$ and $\tau_{a}=\alpha_{a}\beta_{m}+\beta_{a}.$ The above
model is an AFT model since $\widetilde{\varepsilon}$ is independent of
$A\ $ and $C$ which follows from $\left(  \xi,\varepsilon\right)  $ independent of
$A$ and $C$. However, the reduced-form density of $\log T$ given $A$ and $C$ is\ of a
complicated form given by the convolution of a normal density with an extreme
value density: $f_{\widetilde{\varepsilon}}(\cdot) = \int_{\varepsilon}  \frac{1}{\beta_m} f_{\xi}(\frac{\cdot - \sigma \epsilon}{\beta_m}) g(\varepsilon) d\varepsilon$, where $g(\varepsilon)$ is the extreme value density and $\beta_m \neq 0$. Thus, $\widetilde{\varepsilon}$ will not have an extreme value distribution, so that the reduced form model is mis-specified if an extreme value density is assumed for $f_{\widetilde{\varepsilon}}$. As we showed in the previous section, in the presence of censoring, 
the estimator of $\tau_a$ will therefore fail to be consistent; thus, the difference method indirect effect estimator will not be consistent for the indirect effect. 
However, according to our results, in the absence of censoring, the difference method estimator will be consistent for the indirect effect. \\

\noindent\textbf{Equivalence of the product and difference method indirect effect under
	a AFT\ model with a log-normal outcome:} In contrast, if $\varepsilon$ and $\xi$ are both normal, the reduced-form density of $\log T$ given $A$ and $C$ is of correct form because the convolution of two independent normal densities will also be a normal density. Due to this, the reduced form model (\ref{A reduced model}) will be correctly specified, so the estimator of $\tau_a$ will be consistent. Thus, the difference method, $\tau_{a} - \beta_{a}$, will be a consistent estimator for the indirect effect. \\

\noindent\textbf{Monte Carlo variance for indirect effect estimates in the simulation study:}

\renewcommand{\thefigure}{A\arabic{figure}}

\begin{figure}
	\centering
	\caption{Simulation Study, Product vs. Difference Method for the Indirect Effect}
	\includegraphics[scale=.45]{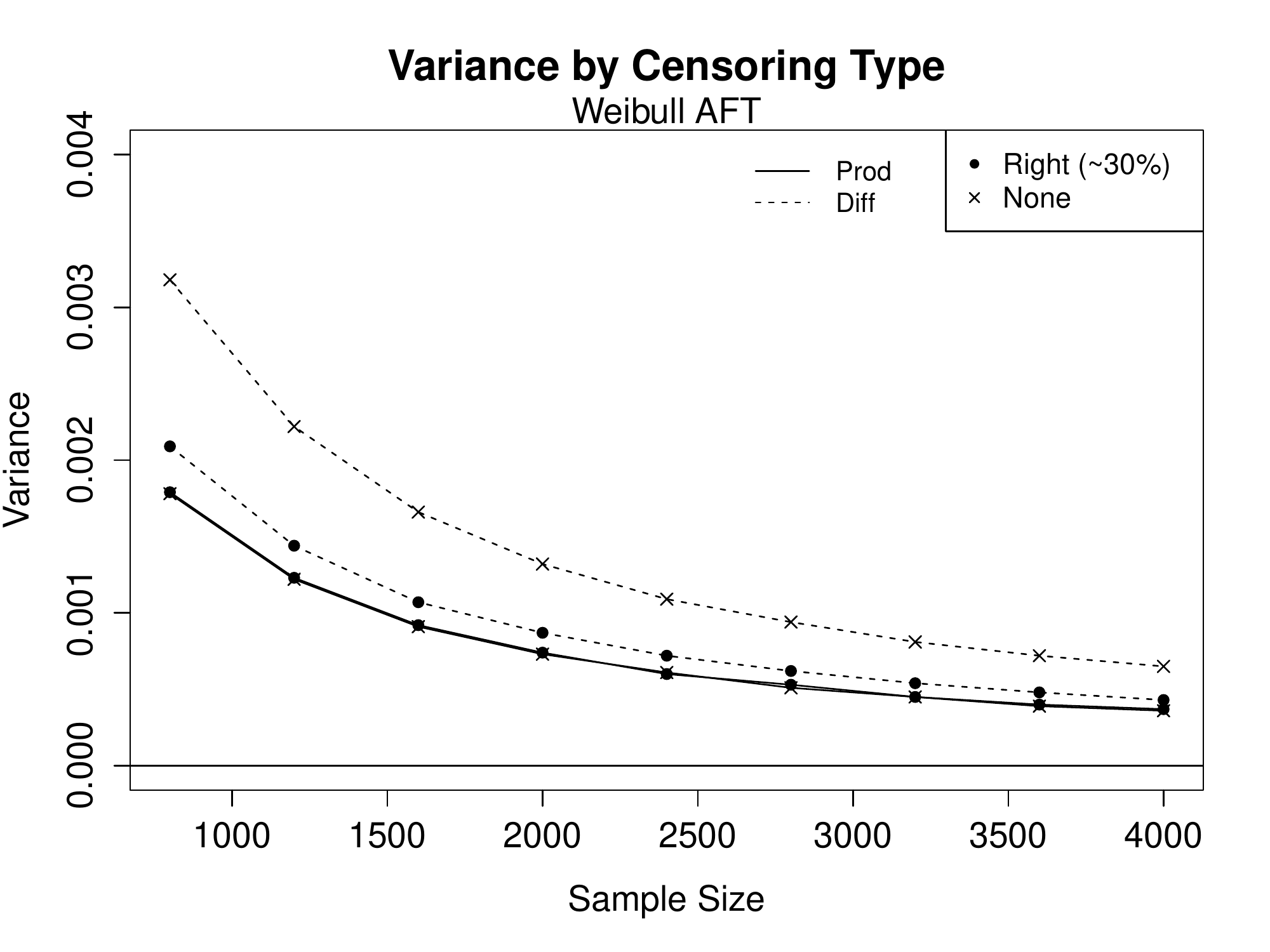}
	\includegraphics[scale=.45]{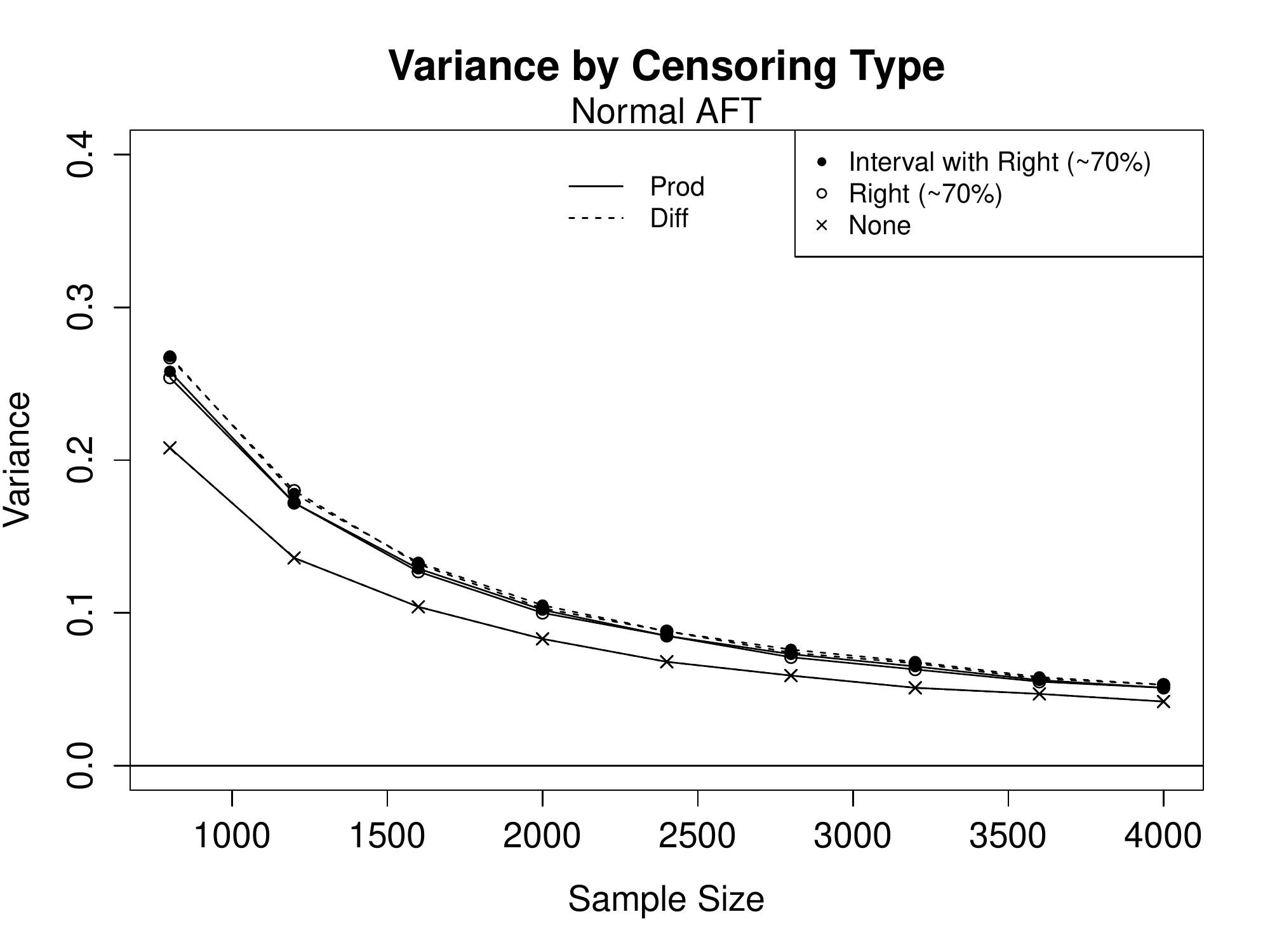}
\end{figure}

\newpage
\noindent\textbf{R Code for direct and indirect effect estimates from data application:}
\begin{verbatim}
##Calculate indirect and direct effect estimates: 
#normally distributed time to event outcome
#normally distributed mediator 
#no interaction between exposure and mediator
#interval and right censoring

#exp is the exposure variable (A in the paper) 
#med is the mediator variable (M in the paper)
#time1 is the left interval 
#time2 is the right interval; NA for right censored data
#cov1,..,cov5 are the potential confounders

#choose the correct library in R
library(survival)

#full model
full.model <- survreg(Surv(time1,time2,type=c('interval2')) ~ exp + med + cov1 + cov2 + 
cov3 + cov4 + cov5, dist="gaussian") 

#reduced model
exp.model <- survreg(Surv(time1,time2,type=c('interval2')) ~ exp + cov1 + cov2 + cov3 
cov4 + cov5, dist="gaussian") 

#mediator model
med.model <- lm(med ~ exp + cov1 + cov2 + cov3 + cov4 + cov5)

#Calculating direct and indirect effects
nde <- full.model$coefficients[2]
nie.prod <- med.model$coefficients[2]*full.model$coefficients[3]
nie.diff <- exp.model$coefficients[2]-full.model$coefficients[2]

#Calculating standard errors for the indirect (product) and direct effect estimates
se_nde <- sqrt(full.model$var[2,2])
se_nie.prod <- sqrt((med.model$coefficients[2]^2)*full.model$var[3,3] + 
(full.model$coefficients[3]^2)*summary(med.model)$cov[2,2])

#################################### 

##Calculate indirect and direct effect estimates: 
#Weibull distributed time to event outcome
#normally distributed mediator 
#no interaction between exposure and mediator
#right censoring

#exp is the exposure variable (A in the paper) 
#med is the mediator variable (M in the paper)
#outcome is the time of event or censoring
#censor is a binary variable indicating censoring
#cov1,..,cov3 are the potential confounders

#full model
full.model <- survreg(Surv(outcome, censor) ~ exp + med + cov1 + cov2 + cov3,      
dist="weibull") 

#reduced model -- Recall the total effect is biased!
exp.model <- survreg(Surv(outcome, censor) ~ exp + cov1 + cov2 + cov3, dist="weibull") 

#mediator model
med.model <- lm(med ~ exp + cov1 + cov2 + cov3)

#Calculating direct and indirect effects
nde <- full.model$coefficients[2]
nie.prod <- med.model$coefficients[2]*full.model$coefficients[3]
nie.diff <- exp.model$coefficients[2]-full.model$coefficients[2] #this is biased!

#Calculating standard errors for the indirect (product) and direct effect estimates
se_nde <- sqrt(full.model$var[2,2])
se_nie.prod <- sqrt((med.model$coefficients[2]^2)*full.model$var[3,3] + 
(full.model$coefficients[3]^2)*summary(med.model)$cov[2,2])


##### NOTES #####
#The nie.diff estimator under the Weibull model will be biased in the presence of censoring
#To calculate the standard errors for the indirect effect (difference), use the boostrap
#Bootstrap code available upon request 



\end{verbatim}

\end{document}